\documentclass[useAMS,usenatbib]{mn2e} \usepackage{mathrsfs}

\usepackage{psfig}
\usepackage{graphicx}
\usepackage{psfig}
\usepackage{amsmath}
\usepackage{threeparttable}
\usepackage{multirow}
\usepackage{booktabs}

\newcommand{\WMAP}{\textit{WMAP}}
\newcommand{\fnl}{$f_{\rm NL}$}
\newcommand{\vecn}{\textbf{\textit{n}}}
\newcommand{\vecr}{\textbf{\textit{r}}}
\newcommand{\vecx}{\textbf{\textit{x}}}

\title[$f^{\rm local}_{\rm NL}$ estimation by CMB skeleton]
{The primordial non-Gaussianity of local type ($f^{\rm local}_{\rm
NL}$) in the \WMAP\ 5-year data: the length distribution of CMB
skeleton}

\author[Hou et al.]
{Zhen Hou$^{1,2,3,4}\footnote{E-mail:houzhen@pmo.ac.cn}$, A. J. Banday$^{5,2}$,
Krzysztof M. G\'orski$^{6,7,8}$, Franz Elsner$^{2}$, \and Benjamin D. Wandelt$^{9,10}$\\
$^{1}$ Purple Mountain Observatory, Chinese Academy of Sciences, 210008, Nanjing, China\\
$^{2}$ Max-Planck-Institute for Astrophysics,
Karl-Schwarzschildstrasse 1, D-85741, Garching bei M\"{u}nchen, Germany\\
$^{3}$ Joint Center for Paritcle, Nuclear Physics and Cosmology, Purple Mountain Observatory
-- Nanjing University, 210093, Nanjing, China \\
$^{4}$ Graduate University of Chinese Academy of Sciences, 100049,
Beijing, China\\
$^{5}$ Centre d'Etude Spatiale des Rayonnements, 9 av du Colonel Roche, BP 44346, 31028 Toulouse Cedex 4, France\\
$^{6}$ Jet Propulsion Laboratory, 4800 Oak Grove Drive, Pasadena CA 91109, USA\\
$^{7}$ California Institute of Technology, Pasadena, CA 91125, USA\\
$^{8}$ Warsaw University Observatory, Aleje Ujazdowskie 4, 00-478 Warszawa, Poland\\
$^{9}$ Department of Physics, University of Illinois
at Urbana-Champaign, 1101 W. Green Street, Urbana, IL 61801-3080, USA\\
$^{10}$ Department of Astronomy, University of Illinois at
Urbana-Champaign, 1002 W. Green Street
Urbana, IL 61801, USA\\
}

\pagerange{\pageref{firstpage}--\pageref{lastpage}}

\pubyear{2010}
\begin{document}

\maketitle

\label{firstpage}

\begin{abstract}
We present skeleton studies of non-Gaussianity in the Cosmic Microwave
Background temperature anisotropy observed in the five-year
\textit{Wilkinson Microwave Anisotropy Probe} (\WMAP) data. The
local skeleton is traced on the 2D sphere by cubic spline
interpolation which leads to more accurate estimation of the
intersection positions between the skeleton and the secondary pixels
than conventional linear interpolation. We demonstrate that the
skeleton-based estimator of non-Gaussianity of the
local type ($f_{\rm NL}^{\rm local}$) - the departure of the length
distribution from the corresponding Gaussian expectation - yields an
unbiased and sufficiently converged likelihood function for $f_{\rm
NL}^{\rm local}$.

We analyse the skeleton statistics in the \WMAP\ 5-year combined V- and
W-band data outside
the Galactic base-mask determined from the KQ75 sky-coverage. The results are
consistent with Gaussian simulations of the the best-fitting
cosmological model, but deviate from the previous
results determined using the  \WMAP\ 1-year data. We show that it is unlikely that the improved
skeleton tracing method, the omission of Q-band data, the modification
of the foreground-template fitting method or the absence of 6 extended regions in the new mask
contribute to such a deviation. However, the application of the Kp0 base-mask in data
processing does improve the consistency with the \WMAP1 results.

The $f_{\rm NL}^{\rm local}$-likelihood functions of the data are
estimated at 9 different smoothing levels. It is unexpected that the
best-fit values show positive correlation with the smoothing
scales. Further investigation argues against a point-source or
goodness-of-fit explanation but finds that about 30\% of either
Gaussian or \fnl\ samples having better goodness-of-fit than the
\WMAP\ 5-year data
show a similar correlation. We present the estimate $f_{\rm NL}^{\rm
local}=47.3\pm34.9$ ($1\sigma$ error) determined from the
first four smoothing angles and $f_{\rm NL}^{\rm local}=76.8\pm43.1$ for the
combination of all nine. The
former result may be overestimated at the
$0.21\sigma$-level because of point sources.
\end{abstract}

\footnotetext{E-mail: houzhen@pmo.ac.cn}

\section{Introduction}
\label{sec_intro}

Generic inflationary models predict that the initial conditions
of the post-inflation universe can be described by a Gaussian
random-phase field with nearly scale-invariant fluctuations.
These subsequently seed the perturbations that generate Cosmic
Microwave Background (CMB) anisotropies and structure formation thereafter.
The Gaussianity of the statistics determined from measures
of the CMB anisotropy and large scale structure distribution can
provide evidence that validates the inflationary scenario of the
extremely-early Universe. Besides the simplest single-scalar field
model that predicts a truly Gaussian initial condition \citep{Guth_1981,
Bardeen_etal_1983, Mukhanov_etal_1992}, there are a number of
inflationary models predicting non-Gaussianity in two broad
classifications, the \textit{equilateral} type and the
\textit{local} type. The detection of a specific type of
non-Gaussianity can
shed light on the fundamental physical properties of inflation.

In this paper, we are concerned with a \textit{local} type non-Gaussianity of
the ``simplest weak nonlinear coupling'' case
\citep{Komatsu_etal_2001}
\begin{equation}
\Phi(\vecx) = \Phi_{\rm L}(\vecx) + f_{\rm NL}^{\rm
local}\left[\Phi_{\rm L}^2(\vecx)-\langle\Phi_{\rm
L}^2(\vecx)\rangle\right],
\end{equation}
where $\Phi(\vecx)$ denotes the primordial curvature perturbations
and $\Phi_{\rm L}$ is its linear Gaussian part. The amplitude of the
non-Gaussianity is parameterised by the dimensionless coupling
constant $f_{\rm NL}^{\rm local}$ (\fnl\ hereafter). The first
observational constraint on \fnl\ -
$-3500 < f_{\rm NL} < 2000$ at 95 \% C.L. - 
was discussed by
\citet{Komatsu_etal_2002} using the angular bispectrum computed from the
four-year \textit{COBE} DMR data \citep{Bennett_etal_1996}.
A reduced bispectrum technique, hereafter the KSW estimator,
\citep{Komatsu_etal_2005}, was applied to the first-year and
three-year \WMAP\ data, leading to $-58 < f_{\rm NL} < 134$
\citep{Komatsu_etal_2003} and $-54 < f_{\rm NL} < 114$
\citep{Spergel_etal_2007}, respectively. \citet{Yadav_Wandelt_2008}
employed an apparently improved estimator \citep{Yadav_etal_2008} to
obtain $27<f_{\rm NL}<147$ for the
V+W-band data outside the Kp0 mask with $\ell_{\rm max}=750$ with the  three-year
\WMAP\ data.
The \WMAP\
team used the same estimator to measure \fnl\ from the five-year \WMAP\
V+W-band outside the KQ75 mask with $\ell_{\rm
max}=700$ and obtained $-9<f_{\rm NL}<111$.

The possibility of detecting CMB non-Gaussianity using
a group of morphological statistics, - Minkowski functionals (MFs)
\citep{Matsubara_2003, Hikage_etal_2006} - has also been studied.
The departure of
MFs from their Gaussian expectations has been tested to be an unbiased
estimator for \fnl\ and then applied to the \WMAP 3-year Q+V+W combined map
yielding $-70<f_{\rm NL}<91$ at the 95\% C.L.
\citep{Hikage_etal_2008}. The \WMAP\ team re-investigated the MFs
estimator with the 5-year template-cleaned V+W map outside the KQ75
mask, yielding $f_{\rm NL}=-57\pm60$ (68\% C.L.) at resolution
$N_{\rm side}=128$ and $f_{\rm NL}=-68\pm69$ at $N_{\rm side}=64$
\citep{Komatsu_etal_2009}. It is still unclear why the MFs favour a
negative best-fit amplitude for \fnl\ while the bispectrum estimator
prefers a positive one, even though  the MFs can be formed by the
weighted sum of the bispectrum. Thus it is of great importance to
use different estimators to identify and investigate the weak
non-Gaussian signal in \WMAP\ observations.
In fact, the one-point probability density function (1-pdf) of the smoothed temperature field
can also be implemented \citep{Bernardeau_etal_2002}
as an alternative non-Gaussianity
estimator \citep{Jeong_etal_2007}. Indeed,
as  noted by \citet{Novikov_etal_2006}, the normalised differential length
of the skeleton is closely linked to this quantity, but the skeleton
remains of interest due to its different sensitivity to specific aspects of the
data, eg. the noise distribution. It is likely that a complete
understanding
of the data can only be realised after the application of a wide range
of statistical tests.

The skeleton has been considered as a probe of the filamentary
structures of a 2D or 3D smooth random field. The original
definition of the skeleton is non-local, making the analytical
discussion difficult and the numerical evaluation costly.
\citet{Novikov_etal_2006} first proposed a local approximation that
``the local skeleton is given by the set of points where the
gradient is aligned with the local curvature major axis and where
the second component of the local curvature is negative''. They also
presented a numerical approach to trace the local skeleton and
found an approximate expression for the differential length
distribution of a Gaussian field. As another morphological
statistical test, the method has been applied to both
large-scale structure measures \citep{Sousbie_etal_2006, Sousbie_etal_2008}
and CMB anisotropies \citep{Eriksen_etal_2004}. The latter was performed
on the Q+V+W map of the first-year \WMAP\ data outside a base-mask
that is defined on the Kp0 sky-coverage. Comparing with Gaussian
simulations, the length distribution of the skeleton did not show
significant deviation from the Gaussian predictions. The impact of
non-excluded point sources was found to be small for the statistics concerned.

In parallel to studies of non-Gaussian signal estimators, several
algorithms of simulating non-Gaussian realisations have been
developed. \citet{Komatsu_etal_2003} first simulated the local-type
non-Gaussian component by integrating the spherical harmonics of
$\Phi_{\rm L}^2(\vecx)-V_{x}^{-1}\int\,d^3\vecx \Phi_{\rm
L}^2(\vecx)$ in spherical harmonic space.
Another strategy has been developed in which a pre-computed `filter' encoding
the
correlation properties of Gaussian curvature perturbation multipoles
boosts the computation of high-resolution temperature and
polarisation Gaussian and corresponding non-Gaussian maps
\citep{Liguori_etal_2003, Liguori_etal_2007}. This method was
recently improved by \citet{Elsner_etal_2009}. Such
\fnl\ simulation methods provide the community with powerful
tools to investigate the primordial non-Gaussianity and the impact
of other astrophysical and systematic effects on it.

In this paper, the skeleton length distribution is adopted as
an estimator of the local-type non-Gaussianity. We adopt
the cubic spline interpolation to trace the underlying local
skeleton rather than the conventional linear one to make a more
accurate estimation of the intersection position between the
skeleton and pixel edge. Motivated by MFs studies on \fnl, the
statistical properties of the skeleton length distribution and the
convergence of an \fnl\ estimation methodology are investigated
from the \fnl\ simulations. We then analyse the skeleton statistics
in the five-year release of the \WMAP\ data and compare with both
Gaussian and non-zero \fnl\ samples. The results of the null
Gaussian test are compared with those of \citet{Eriksen_etal_2004} for
the first-year \WMAP\ data, and then
we use the skeleton estimator to compute a likelihood estimate
for \fnl.

This paper is organised as follows. In Section \ref{sec_method}, we
carry out numerical studies on the CMB local skeleton, including the skeleton statistics utilised in our
analysis (Section \ref{subsec_sta}) and the test of unbiasedness and
convergency of \fnl-likelihood led by skeleton estimator from
noise-free \fnl-simulations (Section \ref{subsec_nulltest}).
\ref{subsec_wmap_data} presents an overview of the \WMAP\ data and
the instrumental properties that should be encoded into our
simulations to make an unbiased comparison and parameter estimation.
Section \ref{subsec_analysis} describes the process of computing the
estimator and further analysis from both the observed data and
simulations having  consistent instrumental properties and
sky-coverage. Results are reported in Section \ref{sec_results},
including the analysis and discussion of a Gaussian frequentist test
(Section \ref{subsec_gauss_res}) and \fnl-estimations
(\ref{subsec_fnl_res}). Finally, we present our conclusions in
Section \ref{sec_conclusions}.

\section{Numerical Studies on CMB local skeleton}
\label{sec_method}

According to the approximation made by \citet{Novikov_etal_2006},
the local skeleton on a smooth 2D sphere $\rho(\vecr)$, traces those
points where the gradient of $\rho$ is the eigenvector of the
corresponding Hessian matrix. That is, it satisfies the
characteristic equation
\begin{equation}
    \mathcal{H}\nabla\rho = \lambda \nabla\rho
\label{eq_det}
\end{equation}
with $\lambda$ ($\lambda_1>\lambda_2$; $\lambda_2<0$) the
eigenvalues, where $\mathcal{H}\equiv\partial^2\rho/\partial
r_i\partial r_j$ is the Hessian matrix at position \vecr. Identically
with \citet{Eriksen_etal_2004}, we do not specify the condition of
eigenvalues of the local linear system. In other words, the skeleton
in our analysis is considered as the set of underlying zero-contour lines
of the realisation
\begin{equation}
    \mathcal{S} = \rho_x\rho_y(\rho_{xx}-\rho_{yy})+\rho_{xy}(\rho^2_y-\rho^2_x),
\label{eq_2D_ske}
\end{equation}
where $\rho_i$ and $\rho_{ij}$ denote the first and second
derivatives of $\rho(\vecr)$ in two orthogonal directions, $x$ and
$y$. As for the CMB temperature field $T(\vecn)$, the `skeleton map'
$\mathcal{S}$ is re-expressed as
\begin{equation}
    \mathcal{S} =
T_{;\theta}T_{;\phi}(T_{;\theta\theta}-
T_{;\phi\phi})+T_{;\theta\phi}(T^2_{;\phi }-T^2_{;\theta}),
\label{eq_CMB_ske}
\end{equation}
where the semicolons denote the covariant derivatives and the
definite expression of them can be found in
\citet{Schmalzing_etal_2002}.

The method for tracing the local skeleton  in the HEALPix scheme has been reviewed in detail by
\citet{Eriksen_etal_2004}. In Appendix \ref{sec_app_1}, we seek to
optimise the method by applying the cubic spline interpolation for
estimating the underlying positions of skeleton `knots' on the
pixelised sphere. The resulting skeleton statistics are introduced
and tested for their applicability to non-Gaussian signal detection and \fnl\ estimation.

\subsection{The statistics}
\label{subsec_sta}

In this work, the CMB temperature realisation intended for skeleton
analysis, $T(\vecn)$, is first normalised as,
\begin{equation}
\nu(\vecn) = \frac{T(\vecn)}{\sigma}.
 \label{eqn_renorm}
\end{equation}
The standard deviation $\sigma$ is computed over the valid region of
\emph{each} realisation after application of  an adequate smoothing process
(Section \ref{subsec_analysis}).

We utilise the skeleton length distribution function of the
normalised temperature thresholds $\nu$, as a probe of
non-Gaussianity and to construct an estimator of \fnl. As with any
probability density function, there are two types of distributions
quantifying the skeleton length, the differential pdf
\begin{equation}
    \mathcal{L}_d(\nu) = \frac{1}{L_{\rm tot}}\frac{dL(\nu)}{d\nu}
    \label{eq_Ld_nu}
\end{equation}
and the cumulative one
\begin{equation}
    \mathcal{L}_a(\nu) = \int^{+\infty}_{\nu}\mathcal{L}_d(\nu')\,d\nu',
\end{equation}
where the normalisation factor $L_{\rm tot} =
\int^{+\infty}_{\nu=-\infty}\,dL(\nu)$ is the total length.

These two functions are equivalent and should lead
to consistent results. In the first investigation of the statistical
properties of the skeleton length in the \WMAP\ data
\citep{Eriksen_etal_2004}, the cumulative
form was utilised and compared with the
predictions of a Gaussian model. In our analysis, both the
differential and
cumulative functions are computed.

\subsection{The idealised skeleton-\fnl-test}
\label{subsec_nulltest}

We study the signature of the local-type non-Gaussianity as a function of
\fnl\ on the skeleton length distributions, $\mathcal{L}_d(\nu)$ and
$\mathcal{L}_a(\nu)$. As a necessary precursor to \fnl-estimation, we
establish that our estimators lead to an unbiased and
sufficiently converged \fnl-likelihood by analysing noise-free full-sky
realisations with a non-Gaussian signal component.
The test is based on simulations of the CMB anisotropy as a function
of \fnl. We adopt the algorithm
proposed by \citet{Liguori_etal_2003, Liguori_etal_2007} and
recently improved by \citet{Elsner_etal_2009} to simulate a set of
Gaussian realisations ($a^{\rm G}_{\ell m}$) with
corresponding non-Gaussian components ($a^{\rm NG}_{\ell m}$). The
cosmological parameters adopted for the \fnl\ simulations are those determined
for the \WMAP5 best-fit $\Lambda$CDM model \citep{Komatsu_etal_2009}.
Specifically, the following parameters are adopted: $\Omega_{\Lambda}=0.742$,
$\Omega_{c}h^2=0.1099$, $\Omega_{b}h^2=0.02273$,
$\Delta_{\mathcal{R}}^2(k_0=0.002 \rm Mpc^{-1})=2.41\times10^{-9}$,
$h=0.719$, $n_s=0.963$, and $\tau=0.087$.
There are a total of 2500 simulated \{$a^{\rm G}_{\ell m}$, $a^{\rm
NG}_{\ell m}$\} pairs in this test that include power up to a maximum
multipole $\ell_{\rm max} = 1024$.

Pixelised skymaps with different \fnl\ values are therefore
obtained following the relation
\begin{equation}
T(p,f_{\rm NL}) = \sum^{\ell_{\rm
max}}_{\ell=2}\sum^{\ell}_{m=-\ell}(a^{\rm G}_{\ell m} + f_{\rm
NL}a^{\rm NG}_{\ell m})b_{\ell}Y_{\ell m}(p),
\end{equation}
where $b_{\ell}$ is a Gaussian beam transfer function with
$\rm{FWHM} = 30\arcmin$ and $60\arcmin$ in this test. The first and
second derivatives of the map can be computed by the HEALPix routine
{\em alm2map\_der}.
Using the method discussed in Appendix \ref{sec_app_1}, the
skeleton length distribution $\mathcal{L}(\nu, f_{\rm NL})$ can then
be estimated from the skeleton map. In this process, the normalised
temperature threshold is set to $\nu \in [-4.0,4.0]$ with 25 uniform
bins.

\begin{figure}
\begin{center}
\includegraphics[width=0.48\textwidth, trim=0.5cm 0.5cm 8.5cm 0.5cm]{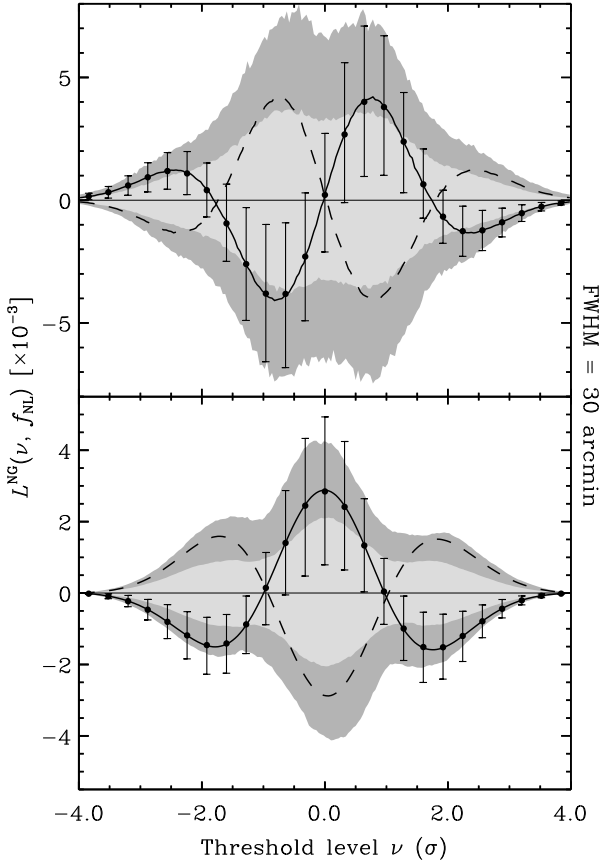}
\end{center}
\caption{The skeleton estimators obtained from 2500 \fnl-simulations
smoothed by a Gaussian beam of $\rm FWHM=30\arcmin$. \textit{Upper}:
The differential estimator $\mathcal{L}^{\rm NG}_{d}$. The solid
curve depicts the averaged distribution, $\mathcal{L}^{\rm
NG}_{d}(\nu,f_{\rm NL}=150)$, of 200 bins, with the error bars
marking the corresponding $1\sigma$ errors of the 25 rebinned values
(black filled circles), while the dashed curve depicts the
estimation of $\langle\mathcal{L}^{\rm NG}_{d}(\nu,f_{\rm
NL}=-150)\rangle$. The grey bands correspond to the
$1\sigma(68.26\%)$ and $2\sigma(95.44\%)$ confidence regions of
$\mathcal{L}^{\rm NG}_{d}(\nu,f_{\rm NL}=0)$, i.e., the Gaussian
condition. \textit{Lower}: The cumulative estimator
$\mathcal{L}^{\rm NG}_{a}$. The nomenclature of elements follows the
same style as the upper panel.} \label{fig_ske_fnl}
\end{figure}

Given the additive nature of the non-Gaussian component, it is
reasonable to express $\mathcal{L}(\nu)$ as
\begin{equation}
\mathcal{L}(\nu, f_{\rm NL}) = \mathcal{L}^{\rm G}(\nu) +
\mathcal{L}^{\rm NG}(\nu, f_{\rm NL}).
\end{equation}
For each $\mathcal{L}(\nu, f_{\rm NL})$ sample, the non-Gaussian
component can be estimated as
\begin{equation}
\mathcal{L}^{\rm NG}(\nu, f_{\rm NL}) = \mathcal{L}(\nu, f_{\rm NL})
- \langle\mathcal{L}^{\rm G}(\nu)\rangle,
\label{eq_ske_NG}
\end{equation}
where $\langle\mathcal{L}^{\rm G}(\nu)\rangle$ gives the Gaussian
expectation of the skeleton length. We depict the samples of
$\mathcal{L}^{\rm NG}(\nu, f_{\rm NL}=0,\pm150)$ in Figure
\ref{fig_ske_fnl}. The grey bands indicate the $1\sigma$ and
$2\sigma$ confidence regions of a purely Gaussian ensemble, $f_{\rm
NL}=0$. It is noteworthy that the behaviour of the non-Gaussian
expectation values
$\langle\mathcal{L}^{\rm NG}(\nu, f_{\rm NL})\rangle$ for both the
differential and cumulative distributions have a characteristic
variation with threshold.
It is similar to MFs in that the peak-trough order
and the amplitude of such features indicate the sign
and the magnitude of \fnl, respectively. This suggests that the skeleton can be
considered as another morphological \fnl-estimator, which may lead
to deeper understanding of the underlying non-Gaussian properties
of the observations. However, with respect to the $1\sigma$ error of
$\mathcal{L}^{\rm NG}$, the fluctuation is roughly within the
$2\sigma$ range of Gaussian predictions, even with $f_{\rm
NL}=150$ which is larger than the  95\% confidence
level upper limit for recent \fnl-estimations using \WMAP\ data. It would still be
challenging for a skeleton estimator to provide a firm
Gaussian/non-Gaussian assessment using the observed data.

Considering only the diagonal elements of the covariance matrix, we use
2000 simulations to estimate $\langle\mathcal{L}^{\rm
G}(\nu)\rangle$, the mean and the standard deviation of
$\mathcal{L}^{\rm NG}(\nu, f_{\rm NL})$. The 500 remaining simulations are
used to compute the $\chi^2$ functions. Given a hypothetical value
of $f^{\rm true}_{\rm NL}$, the $\chi^2(f_{\rm NL}|f^{\rm true}_{\rm
NL})$ of each \fnl-skeleton sample with index $i$ ($i=1,2,...500$)
is computed as
\begin{equation}
\chi^2(f^{i}_{\rm NL}|f^{\rm true}_{\rm NL})  =
\sum_{\nu}\left[\frac{\mathcal{L}^{\rm NG}_{i}(\nu, f_{\rm
NL})-\langle\mathcal{L}^{\rm NG}(\nu, f^{\rm true}_{\rm
NL})\rangle}{\sigma(\mathcal{L}^{\rm NG}(\nu, f^{\rm true}_{\rm
NL}))}\right]^2, \label{eq_chi2}
\end{equation}
where the correlations between bins have not been taken into account
because the full covariance matrix is not sufficiently converged
for the available sample volume in our analysis. Further tests
indicate that the corresponding likelihood from each sample is of
bimodal or even multi-modal shape if the full covariance matrix
is adopted, which causes the estimation to be unrevealing.

\begin{table}
\caption{The best-fit \fnl\ and $1\sigma$ error from the
likelihood $\exp\left[-\frac{1}{2N}\sum^{N}_{i=1}\chi^2(f^{i}_{\rm
NL}|f^{\rm true}_{\rm NL})\right]$ (Figure \ref{fig_likelihood})
computed by differential and cumulative estimators from $N=500$
noise-free simulations with input parameter, $f_{\rm NL}^{\rm
true}=0,\pm150$.}
\begin{center}
    \begin{tabular}{c|c|c|c|c|c}
        \toprule
        estimator & $f^{\rm true}_{\rm NL}$ & \multicolumn{2}{c}{$f^{\rm best}_{\rm
        NL}$} & \multicolumn{2}{c}{$\sigma_{f_{\rm
        NL}}$} \\
        & & $30\arcmin$ & $60\arcmin$ & $30\arcmin$ &
        $60\arcmin$ \\
        \midrule
        \multirow{3}{*}{$\mathcal{L}^{\rm NG}_{\rm d}(\nu,f_{\rm NL})$} & -150.0
        & -151.5 & -148.0 & 26.1 & 37.8 \\
        & 0.0 & 0.8 & 2.1 & 25.3 & 37.5 \\
        & 150.0 & 153.6 & 152.7 & 26.0 & 37.8 \\
        \midrule
        \multirow{3}{*}{$\mathcal{L}^{\rm NG}_{a}(\nu,f_{\rm
        NL})$} & -150.0 & -151.4 & -147.7 & 23.1 & 33.5 \\
        & 0.0 & 0.7 & 1.9 & 22.4 & 33.2 \\
        & 150.0 & 153.2 & 152.0 & 23.1 & 33.5 \\
        \bottomrule
    \end{tabular}
\end{center}
\label{tab_nulltest}
\end{table}

\begin{figure}
\begin{center}
\includegraphics[width=0.48\textwidth, trim=1cm 0.7cm 2cm 1.7cm]{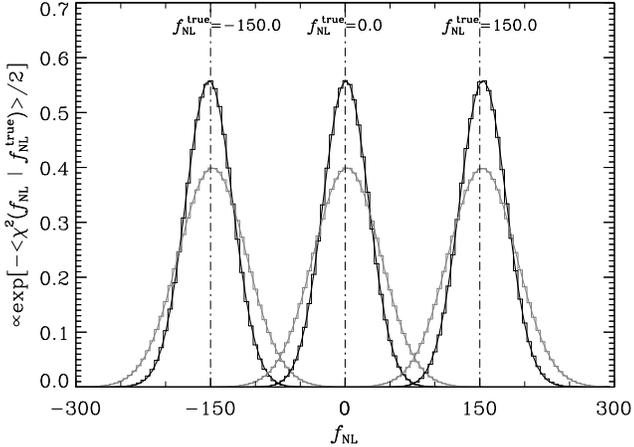}
\end{center}
\caption{$\exp\left[-\frac{1}{2N}\sum^{N}_{i=1}\chi^2(f^{i}_{\rm
NL}|f^{\rm true}_{\rm NL})\right]$, the effective likelihood
functions computed by the differential estimator $\mathcal{L}^{\rm
NG}_{d}$ with input parameter $f_{\rm NL}^{\rm true}=0,\pm150$. The
actual functions are renormalised by different factors for visual
convenience and are shown by histograms. They are extremely well
fitted by the Gaussian functions depicted by solid curves. The higher,
narrower (lower, wider) histograms and curves correspond to
likelihoods from simulations with $\rm FWHM=30\arcmin$
($60\arcmin$).} \label{fig_likelihood}
\end{figure}

The parameter \fnl\ is uniformly sampled from $-300$
to $300$ with a step-length $\Delta f_{\rm NL}=5$. We estimate the
likelihoods for three specific $f_{\rm NL}^{\rm true}$ values, 0 and
$\pm150$. The posterior PDF for $f^{\rm true}_{\rm NL}$ can be
obtained by Bayes' theorem
\begin{equation}
\begin{split}
P(f^{\rm true}_{\rm NL}|\{f^{i}_{\rm NL}\}) & \propto P(\{f^{i}_{\rm
NL}\}|f^{\rm true}_{\rm NL})\times P(f^{\rm
true}_{\rm NL}) \\
& \propto \prod^{N}_{i=1}P(f^{i}_{\rm NL}|f^{\rm true}_{\rm
NL}) \\
& \propto \exp\left[-\frac{1}{2}\sum^{N}_{i=1}\chi^2(f^{i}_{\rm
NL}|f^{\rm true}_{\rm NL})\right],
\end{split}
\end{equation}
where we have conservatively set the prior $P(f^{\rm true}_{\rm NL})$
to be uniform and $N$ equals to 500. In fact,  we
have found that roughly 20 samples with FWHM$=30\arcmin$ smoothing
are adequate for the posterior distribution to converge sharply around
$f^{\rm true}_{\rm NL}$  with a $1\sigma$ error $\simeq
\Delta f_{\rm NL}=5$. However, we have only one observed CMB sample
so that the convergence of the consequent posterior distribution is
limited by the data resolution and the noise level. The effective
likelihood functions of each sample, i.e.,
$\exp\left[-\frac{1}{2N}\sum^{N}_{i=1}\chi^2(f^{i}_{\rm NL}|f^{\rm
true}_{\rm NL})\right]$, are illustrated in Figure
\ref{fig_likelihood}, using different normalisation factors for visual
convenience. The histograms depict the computed likelihoods which
are perfectly fitted by Gaussian functions. Accordingly, the mean
and the $1\sigma$ width of each likelihood are estimated as
presented in Table \ref{tab_nulltest}.

The results demonstrate a good recovery of the input \fnl\ values
given the interval $\Delta f_{\rm NL}=5$ of our sampling. The
$\chi^2(f_{\rm NL}|f^{\rm true}_{\rm NL})$ in Eq. \ref{eq_chi2}
therefore constitutes an unbiased maximum likelihood position in
\fnl-space and the corresponding $1\sigma$ error is determined by
the likelihood function. It is noteworthy that the cumulative
estimator behaves a little bit better than the differential one and
therefore the former is selected for \fnl\ estimation as applied to real
data.

\section{Method}

Even though the literature contains theoretical predictions for the
length distributions of the local skeleton on a 2D Gaussian random field,
our analysis compares measures derived from simulated observations
of the sky with the corresponding values for the \WMAP\ data, since
the inhomogeneous noise contribution and the complicated
sky-coverage render analytical investigation difficult. Furthermore, it
is also difficult to interpret the non-Gaussian component of these
skeleton measures analytically. In what follows, we introduce both the
instrumental properties impacting the observed data and the essential
numerical processing steps required for further analysis.

\subsection{The \WMAP\ data and the simulations}
\label{subsec_wmap_data}

The \WMAP\ instrument measures the CMB temperature anisotropy in
five frequency bands from 23 to 94 GHz \citep{Bennett_etal_2003a}.
The foreground-reduced sky maps in V and W-band are used in our
analysis, identical to the data selection for the \WMAP\ five-year
power spectrum estimation \citep{Nolta_etal_2009}. These maps are
available in the HEALPix pixelisation scheme with $N_{\rm side}=512$
from the LAMBDA
website\footnote{http://lambda.gsfc.nasa.gov/product/map/dr3/\\maps\_da\_forered\_r9\_iqu\_5yr\_get.cfm}.
The maps from two (four) differencing assemblies (DAs) at V- (W-) band
are combined using uniform weights over the sky and equal weights for
each DA. The resulting maps in V- and W-band are then combined to
obtain the VW-band map using the same method. The effective beam
transfer function of the VW-band can then be easily computed from the beam functions
of those
DAs\footnote{http://lambda.gsfc.nasa.gov/product/map/dr3/beam\_info.cfm}
constituting the VW-band map. The
observational data are inevitably affected by the instrumental
noise, dominated by an uncorrelated component with a variance per
pixel depending on the noise
amplitude $\sigma_0$ and the pixel scanning strategies of each DA,
$N_{\rm obs}(p)$ \citep{Bennett_etal_2003a}.

\begin{figure}
\includegraphics[width=0.27\textwidth,angle=90,trim=0.8cm 0 0.8cm 0]{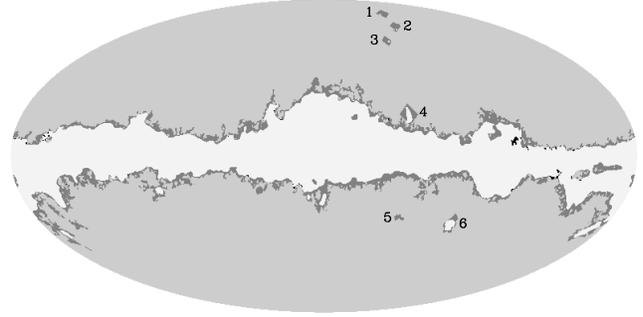}
\caption{Comparison of the KQ75 base-mask (KQ75B) used in \WMAP5 data
processing with the Kp0 base-mask (KQ75B) used for \WMAP1 and \WMAP3
analysis. The white (light-grey)
regions are excluded (included) by both KQ75B and Kp0B.
The dark-grey (black) parts are excluded by the former (latter) but not
excluded by the latter (former). There are 6 extended regions
(labeled from `1' to `6') eliminated in the KQ75B mask.}
\label{fig_base_mask}
\end{figure}

The extended temperature analysis mask (KQ75) is adopted to
minimise the contamination from the diffuse Galactic foreground and
point source emission. For further investigations, the part related to
the Galactic emission
is separated out to form a base-mask called
`KQ75B' in our analysis. As a comparison of the base-masks used
in the 1-, 3- and 5-year \WMAP\ data analyses, we illustrate the
Kp0B (adopted by \citet{Eriksen_etal_2004}) and KQ75B mask in
Figure \ref{fig_base_mask}. Besides the extended Galactic profile,
there are six extended regions (labeled from `1' to `6') eliminated by the new
base-mask. The impact of these regions on
skeleton statistics will be considered when comparing our results with
those of \citet{Eriksen_etal_2004} for \WMAP1.

The 2500 pairs of Gaussian and non-Gaussian realisations $\{a_{\ell
m}^{\rm G}, a_{\ell m}^{\rm NG}\}$ introduced in Section
\ref{subsec_nulltest} are used for our \fnl\ studies. For each \fnl\ value, we construct
a map with resolution parameter $N_{\rm side}=512$ and \WMAP\
instrumental properties as
\begin{eqnarray}
T(p,f_{\rm NL}) & = & \sum^{\ell_{\rm
max}}_{\ell=2}\sum^{\ell}_{m=-\ell}(a^{\rm G}_{\ell m} + f_{\rm
NL}a^{\rm NG}_{\ell m})b_{\ell}p_{\ell}Y_{\ell m}(p) \nonumber \\
& & \mbox{} + \frac{\sigma_0 \eta}{N_{\rm obs}(p)}, \label{eq_fnl_map}
\end{eqnarray}
where $b_{\ell}$ is the effective beam transfer function of the \WMAP\
VW-band data and $p_{\ell}$ is the pixelisation window function for
$N_{\rm side}=512$. The second term on the \textit{rhs} simulates
the noise contribution on each pixel with Gaussian random number $\eta \sim N(0,1)$.

In this work, we perform both a Gaussian frequentist test and
\fnl-estimations. In the former, the Gaussian simulations are
processed in the same way as Eq. \ref{eq_fnl_map} but free of the
\fnl\ term.

\subsection{Data processing and the analysis}
\label{subsec_analysis}

In this section, we introduce the data processing methods applied to
both the observed and the simulated realisations for studies of the
skeleton length distribution. The processing steps presented here
follow the strategy detailed in Section 4 of
\citet{Eriksen_etal_2004}.

\subsubsection{Map processing}

The base-mask is applied to the map to avoid Galactic foreground
contamination. Following the methodology of \citet{Eriksen_etal_2004},
we do not exclude point sources,
in particular because any additional smoothing applied
to the mask reduces the sky coverage available for analysis dramatically.
This approach is supported by studies 
of the spectral parameter, $\gamma$, by \citet{Eriksen_etal_2004},
which indicates that smoothing of the data renders the skeleton
less sensitive to point source signal contributions for larger
FWHMs. Moreover, a median-filter technique is applied to the point
sources to investigate their impact on the skeleton statistics for
smaller smoothing FWHMs.
Specifically, for a given pixel $i$ that would be eliminated by the
point-source mask, we consider all other unmasked pixels
within a $1^\circ$ radius and determine the median temperature
for this set of pixels. The temperature at pixel $i$ is then replaced
by this median value, and the process repeated for all pixels specified
in the point-source mask.
The median-filtered map is then analysed in the same manner as the
unfiltered data set.

Following standard procedure in CMB data analysis, the monopole and
dipole components are fitted and removed from each map outside the masked
region. This step is achieved using the HEALPix F90 subroutine, {\em remove\_dipole}. The resulting map is then smoothed with a Gaussian beam in harmonic space, again using HEALPix tools.
For easy comparison with the previous analysis of
\citet{Eriksen_etal_2004},
the FWHM widths, $\theta_{\rm FWHM}$, selected for these Gaussian smoothing beams
are taken to be
$0\fdg53$, $0\fdg64$, $0\fdg85$,
$1\fdg28$, $1\fdg70$, $2\fdg13$, $2\fdg55$, $2\fdg98$ and $3\fdg40$
(we abandon the larger angular scales used in the former analysis to ensure good convergence of the
\fnl-likelihood).
Since a higher resolution map is necessary for more
accurate estimation of the skeleton statistics (see Appendix
\ref{sec_app_1}), the resolution parameter of the resulting smoothed
map is set to $N_{\rm side}=1024$.

After the smoothing of the data, the base-mask must also be
expanded and the same processing method is followed.
For each smoothing scale, only those base-mask pixels
with values larger than 0.99 are defined to be valid
pixels on the smoothed mask.

Finally, on the valid region defined by the smoothed base-mask, the map $\tilde T(p)$,
either from observation or simulation, is renormalised to
temperature thresholds, $\nu$ (Eq.\ref{eqn_renorm}), while the
invalid pixels are abandoned for computing the standard deviation.
Using the method discussed in Appendix \ref{sec_app_1}, the
skeleton length distributions, $\mathcal{L}_{d}(\nu)$ and
$\mathcal{L}_{a}(\nu)$, can be estimated for each set of
smoothed samples. The original distribution $L(\nu)$ in Eq.
\ref{eq_Ld_nu} is divided into 200 bins with $\nu \in [-4.0, 4.0]$
during skeleton tracing.

\subsubsection{Non-Gaussian detector and estimator}
\label{subsubsec_estimator}

From the processed Gaussian simulations, we compute the
Gaussian expectation of the skeleton statistics for each smoothing
scale, $\langle \mathcal{L}^{\rm G}(\nu, \theta_{\rm FWHM})
\rangle$.
The departure from these expectation values is then obtained
for both the observed data and each Gaussian sample as
\begin{equation}
\Delta \mathcal{L}(\nu,\theta_{\rm FWHM}) =
\mathcal{L}(\nu,\theta_{\rm FWHM}) - \langle \mathcal{L}^{\rm
G}(\nu, \theta_{\rm FWHM}) \rangle,
\end{equation}
and the corresponding $\chi^2$ value is then computed
\begin{equation}
\chi^2 = \left[\frac{\Delta \mathcal{L}(\nu,\theta_{\rm
FWHM})}{\sigma(\Delta \mathcal{L}(\nu,\theta_{\rm FWHM}))}\right]^2
\end{equation}
where we omit the $\langle\Delta \mathcal{L}\rangle$ term since it
is definitely zero. In the \fnl\ analysis, the non-Gaussian departure,
$\mathcal{L}^{\rm NG}(\nu,f_{\rm NL},\theta_{\rm FWHM})$, and the
$\chi^2$ statistics are estimated by Eq.\ref{eq_ske_NG} and
\ref{eq_chi2}. The best-fit value and error of \fnl\ can then be
obtained by analysing the likelihood function as discussed in Section
\ref{subsec_nulltest}.

Before we provide final estimates of \fnl\ from the different
smoothing scales, we combined the estimators, $\Delta
\mathcal{L}(\nu)$ of the data and $\mathcal{L}^{\rm NG}(\nu, f_{\rm
NL})$ of each set of \fnl\ sample, to
\begin{equation}
\Delta \mathcal{L}_{\rm C}(\nu,f_{\rm NL}) = \sum_{i=1}^{N_{\rm
fwhm}}w_i(\nu,f_{\rm NL}) \Delta \mathcal{L}^{i}(\nu)
\end{equation}
and
\begin{equation}
    \mathcal{L}^{\rm NG}_{\rm C}(\nu,f_{\rm
NL}) = \sum_{i=1}^{N_{\rm fwhm}}w_i(\nu,f_{\rm NL})
\mathcal{L}^{{\rm NG},i}(\nu,f_{\rm NL})
\end{equation}
respectively with the inverse-variance weighting
\begin{equation}
w_i(\nu,f_{\rm NL}) = \frac{1/\sigma^2(\mathcal{L}^{{\rm
NG,}i}(\nu,f_{\rm NL}))}{\sum^{N_{\rm fwhm}}_{i=1}
1/\sigma^2(\mathcal{L}^{{\rm NG},i}(\nu,f_{\rm NL}))}
\end{equation}
where $i$ corresponds to one smoothing scale and $N_{\rm fwhm}$
represents the number of scales used in the combination. The combined
$\chi^2$ is then computed
\begin{equation}
    \chi^2_{\rm C}(f_{\rm NL})=\sum_{\nu} \left\{ \frac{\Delta \mathcal{L}_{\rm C}(\nu,f_{\rm
    NL})-\langle \mathcal{L}^{\rm NG}_{\rm C}(\nu,f_{\rm
NL}) \rangle}{\sigma[\mathcal{L}^{\rm NG}_{\rm C}(\nu,f_{\rm NL})]}
\right\}^2.
\end{equation}
This combination makes an integrated estimation of \fnl\ which
includes the non-Gaussian signal at several different scales with a
mild weighting.

\section{Results and discussions}
\label{sec_results}

\subsection{Gaussian frequentist results}
\label{subsec_gauss_res}

We first compare the observed results with our Gaussian model predictions.
In this case, we perform 10240 Gaussian simulations of the \WMAP\
VW-band properties. Different base-masks, as well as the
median-filter, are applied independently to both the real and the simulated skies
to study the foreground effect on the skeleton
results. The corresponding $\chi^2$ values are then computed to
enable the frequentist test.

\subsubsection{Results of KQ75B processing}

\begin{figure}
    \includegraphics[angle=90, width=0.44\textwidth, trim=0 12.3cm 0.cm
0]{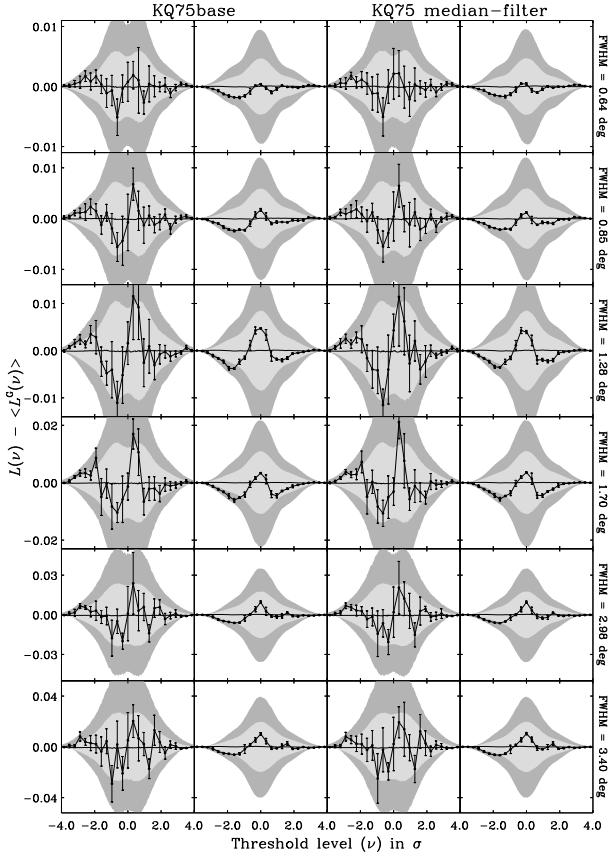} \caption{The skeleton statistics, $\Delta
\mathcal{L}(\nu)=\mathcal{L}(\nu)-\langle \mathcal{L}^{\rm G}(\nu)
\rangle$, computed from KQ75B (left two columns) and KQ75
median-filter (right two columns) processing on smoothing scales
$\rm FWHM=0\fdg64$, $0\fdg85$, $1\fdg28$, $1\fdg70$, $2\fdg98$ and
$3\fdg40$. The first and third column (the second and forth column)
correspond to results of the differential (cumulative) estimator.
The grey bands show the $1\sigma$ and $2\sigma$ confidence regions
defined by 10240 Gaussian samples. The black filled-circles
connected by solid lines show the observed sample which is rebinned
to 25 bins. The error bar mark the $1\sigma$ error in each bin
according to such rebinning.} \label{fig_dske_KQ75}
\end{figure}

For each smoothing scale, the skeleton length departure from the
Gaussian expectation, $\Delta \mathcal{L}(\nu,\theta_{\rm FWHM}) =
\mathcal{L}(\nu,\theta_{\rm FWHM})-\langle \mathcal{L}^{\rm
G}(\nu,\theta_{\rm FWHM}) \rangle$, is computed from samples
obtained with the KQ75B masked maps. The results are shown in the
left two columns (for both the differential and cumulative
distributions) of Figure \ref{fig_dske_KQ75} for $\theta_{\rm FWHM}
= 0\fdg64$, $0\fdg85$, $1\fdg28$, $1\fdg70$, $2\fdg98$ and
$3\fdg40$. The grey bands demonstrate the $1\sigma$
and $2\sigma$ confidence regions of the Gaussian prediction. The
observed ones are rebinned to 25 bins and depicted by filled circles
with the $1\sigma$-error bar of each bin. The rebinning is necessary
since the differential skeleton distribution is relatively noisy.

\begin{figure*}
    \begin{center}
        \includegraphics[width=0.85\textwidth, trim=0 -0.5cm 2.0cm 0]{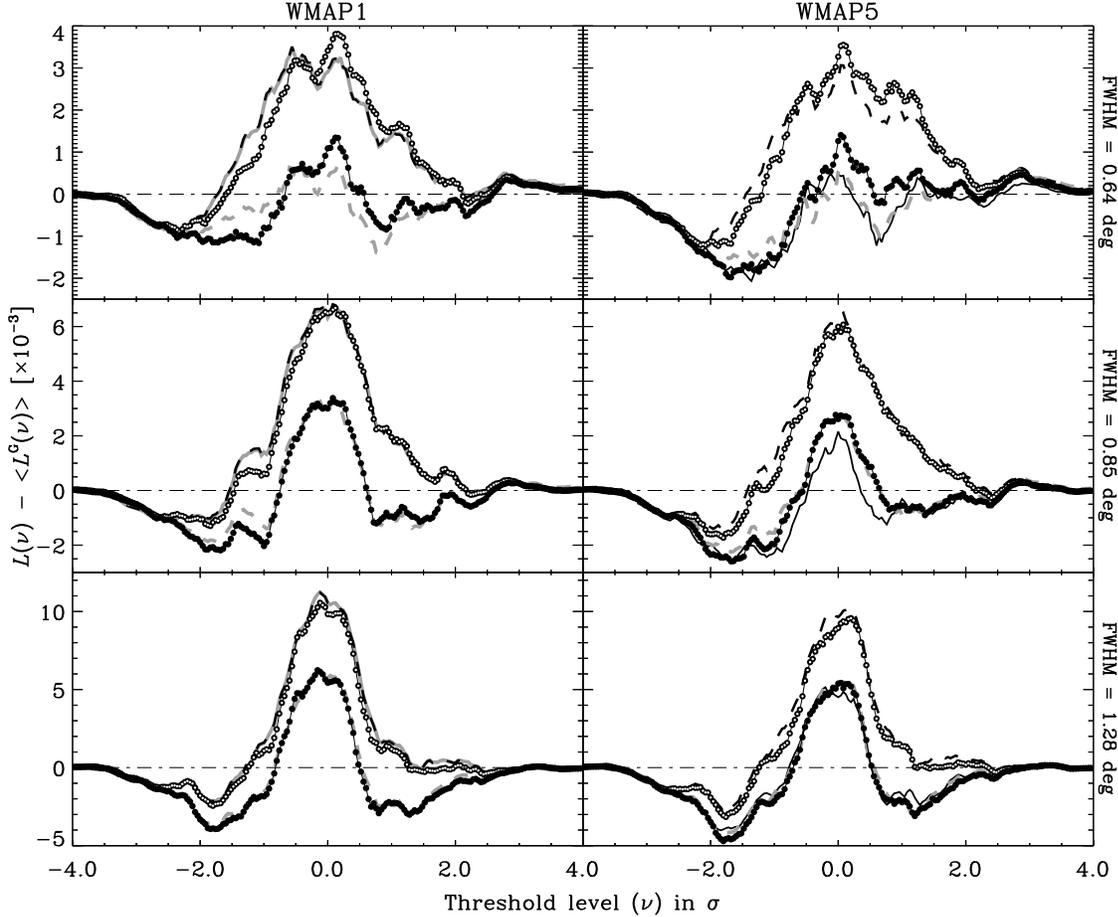}
        \caption{The skeleton statistics, $\Delta \mathcal{L}_{a}(\nu)$, computed from different processing methods. \emph{Left Column}: the results obtained from the one-year \WMAP\ data. \emph{Right Column}: the results obtained from the five-year \WMAP\ data. The lines and symbols denoting different processing elements are noted in Table \ref{tab_sys_colo}.} \label{fig_syseff}
    \end{center}
\end{figure*}

In the case of the cumulative distributions, $\Delta
\mathcal{L}_{a}(\nu)$ for \WMAP5, some features consistent with a positive
\fnl\ value are observed, albeit within the
$1\sigma$ Gaussian confidence level. The behaviour of the differential
distribution, $\Delta \mathcal{L}_{d}(\nu)$, supports this
inference despite the existence of a higher level of fluctuations.
However, there are
differences between the new results and the corresponding \WMAP1 ones
\citep{Eriksen_etal_2004}. For each smoothing scale, the latter show
a $1\sigma$-level peak around $\nu=0$ while the neighbouring troughs
show less fluctuations especially in the $\nu>1$ region. In contrast, as
shown in Figure \ref{fig_dske_KQ75} (the left two columns), the
former's peak is less apparent but the troughs are much more
distinct particularly for $\theta_{\rm FWHM}=1\fdg28$ and $1\fdg70$. The
comparison between \WMAP1 and our new results is shown in Figure
\ref{fig_syseff} for $\theta_{\rm
FWHM}=0\fdg64$, $0\fdg85$ and $1\fdg28$.

There are several possibilities associated with such a discrepancy.

\begin{enumerate}
\renewcommand{\theenumi}{(\arabic{enumi})}

\item Change of the skeleton-tracing method.
    Utilising cubic spline interpolation in the skeleton tracing
  algorithm yields a more accurate estimation of the quantities than
  the previously adopted linear algorithm (see Appendix
  \ref{sec_app_2}).  We computed $\Delta \mathcal{L}_{a}(\nu)$ for the
  template-cleaned \WMAP1 data using the same band-selection, mask and
  processing steps as in \citet{Eriksen_etal_2004}, and tracing the
  underlying skeleton by both linear and cubic spline interpolation
  strategies. The Gaussian expectation is also estimated in both cases
  using simulations. The results are shown in the left column of
  Figure \ref{fig_syseff} where it can be seen that the dashed black line (cubic spline) and the solid grey line (linear) essentially overlap. It
  is therefore clear that changing the interpolation scheme
  contributes little to the discrepancy found. This issue is also
  discussed in Appendix \ref{sec_app_2} for the \WMAP5 data.

\item Band-selection.
    In the analysis of \citet{Eriksen_etal_2004}, the Q-, V- and W-band
  maps are combined with a spatially-invariant inverse-noise-variance
  weigthing. The resulting map is dominated by the Q-band since it has
  the lowest noise of the three. However, since it is the band for
  which Galactic foreground residuals remain significant, it is
  plausible that these have an impact on the skeleton results.
  We repeated our analysis using the appropriately weighted \WMAP5 Q-,
  V- and W-band data, but retaining the KQ75B base-mask.
  Corresponding Gaussian simulations are also performed.  The results
  are shown as the black connected-filled-circles in the right column of
  Figure \ref{fig_syseff}. The profile shows modest deviation from our
  VW-results (black filled-squares), however, it does not result in the discrepancy
  level required. On the contrary, the difference becomes less
  significant for large $\theta_{\rm FWHM}$.

\item Difference of the foreground subtraction method between \WMAP1 and \WMAP5.
    The foreground templates used for the former
  \citep{Bennett_etal_2003b} are the FDS 94 GHz dust prediction, the
  H$\alpha$ map for free-free emission and the 408 MHz Haslam map for
  synchrotron emission. The three-year \WMAP\ foreground analysis
  \citep{Hinshaw_etal_2007} and beyond replace the 408 MHz data with a
  template based on the the K-Ka difference map.  The difference
  between the two foreground models at V-band utilising the
  coefficients for the first-year fits of \citet{Bennett_etal_2003b}
  and the five-year analysis of \citet{Gold_etal_2009} is shown in
  Figure \ref{fig_fore_diff}. The profile demonstrates a dipole-like
  structure in the large-scale temperature distribution outside both
  the Kp0B or KQ75B masks, which may affect the skeleton
  statistics and the corresponding inferences of \fnl.  We subtract
  the five-year foreground model from the one-year raw maps at Q-, V-
  and W-bands, which are then combined and processed identically with
  \citet{Eriksen_etal_2004} using the Kp0B mask. The corresponding
  skeleton statistic, $\Delta \mathcal{L}_{a}(\nu)$, is depicted by
  the connected open-circles in the \emph{left} column of Figure
  \ref{fig_syseff}.  They demonstrate consistency with the original
  \WMAP1 results. Similarly, another independent test has been carried
  out on the five-year raw maps from which the one-year foreground
  model is subtracted before the data are combined and processed using
  the KQ75B mask. The results are depicted as the dashed grey line in the \emph{right} column of Figure
  \ref{fig_syseff}, and demonstrate consistency with our five-year
  templated-cleaned VW-KQ75B results (black filled-squares). We
  conclude that it is difficult to attribute the observed discrepancy
  to the change of foreground subtraction method.

\begin{figure}
    \includegraphics[width=0.48\textwidth]{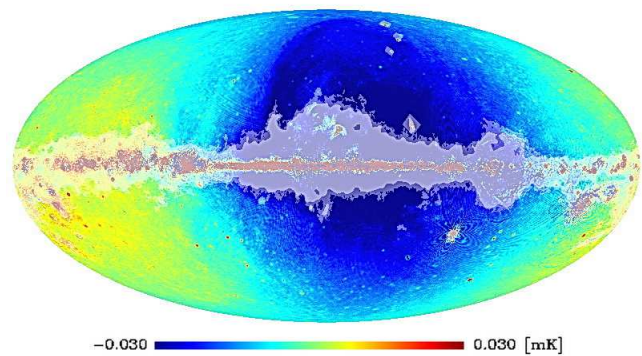}
\caption{The difference map between the \WMAP1 and \WMAP5 combined foreground
models at V-band. between \WMAP1 and \WMAP5. The Kp0B
and KQ75B masks are also denoted. The ringing effect around some bright
sources, especially the LMC, comes from the one-year processing of the
three templates \citep{Bennett_etal_2003b}.}
\label{fig_fore_diff}
\end{figure}

\begin{table}
\caption{The processing elements for the styles of lines and symbols in
Figure \ref{fig_syseff}.}
\centering
\begin{threeparttable}
    \scriptsize{
    \begin{tabular}{c|c|c|c|c}
        \toprule
        Band & Mask\tnote{a} & Fore-red\tnote{b} & Interp\tnote{c} & Style \\
        QVW  & Kp0B  & 1yr & Cubic  & Dashed black line \\
        QVW  & Kp0B  & 5yr & Cubic  & Connected open-circles \\
        QVW  & KQ75B & 1yr & Cubic  & Dashed grey line \\
        QVW  & KQ75B & 5yr & Cubic  & Connected filled-circles \\
        QVW  & Kp0B  & 1yr & Linear & Solid grey line \\
         VW\tnote{d}  & KQ75B & 5yr & Cubic  & Solid black line \\
        \bottomrule
    \end{tabular}}

    \footnotesize
    \begin{tablenotes}
        \item [a] The base-mask applied in map-processing and analysis.
        \item [b] The templates and the corresponding coefficients applied for
foreground-reducing before our map-processing.
        \item [c] The interpolation method used for tracing the underlying
local skeleton.
        \item [d] The V+W combined data with uniform weighting, while spatial
invariant inverse-noise-variance weighting for QVW.
    \end{tablenotes}
\end{threeparttable}

\label{tab_sys_colo}
\end{table}

\item Change of the base-mask in processing.
    It is very suspicious that the residual foreground components around
  the dark-grey regions in Figure \ref{fig_base_mask} bias the skeleton
  results of \WMAP1, although mild smoothing and mask thresholding are
  applied before skeleton tracing. We discuss this issue in Section
  \ref{subsubsec_kqp0} by investigating the Galactic plane region and
  the extragalactic sources (labeled from 1--6 in Figure
  \ref{fig_base_mask}) separately.

\end{enumerate}

As shown by the left two columns in Figure \ref{fig_dske_KQ75}, the
profile of the \WMAP5 $\Delta \mathcal{L}_{a}(\nu)$ function is
consistent with that expected for a positive-\fnl. In particular, both $\Delta
\mathcal{L}_{d}$ and $\Delta \mathcal{L}_{a}$, rebinned for
$\rm{FWHM}=1\fdg28$ and $1\fdg70$, demonstrate consistent features
with the solid lines shown in Figure \ref{fig_ske_fnl} for $f_{\rm
NL}= +150$. However, the troughs in the $\nu>1$ region (hot
region) seem relatively less depressed. It is likely that the point
sources and foreground components contribute to this asymmetry
between the two troughs. The results from the median-filtered map
yield insights implications into this issue.

We computed the $\chi^2$ values of $\Delta \mathcal{L}_{a}$ for both
the observed and the simulated samples.  We list the fraction of
the simulations with a $\chi^2$ values less extreme than the observed one
in Table \ref{tab_Gfreq}. The corresponding
\WMAP1 results are also listed (Table 3 in
\citet{Eriksen_etal_2004}). Generally speaking, there is no
qualitative difference between the five-year and one-year
results. But our results show a unimodal
dependence on the smoothing scales.
The \fnl-signal seems more significant around the angular scales
$\rm{FWHM}=1\fdg28$, $1\fdg70$ and $2\fdg13$.

\begin{table}
\caption{The $\chi^2$-based frequentist results for the \WMAP5 skeleton analysis
derived using different processing masks and methods on 10 smoothing
scales.  We list the fraction of
the simulations with a $\chi^2$ values less extreme than the observed one.
The letters `M' correspond to `median-filter'. The values are determined from the estimator
$\Delta \mathcal{L}_{a}(\nu)$ computed for 200 bins from the data and 10240
Gaussian samples. The corresponding results for \WMAP1
\citep{Eriksen_etal_2004} are also listed for easy comparison.}
\begin{center}
    \begin{tabular}{c|c|c|c|c|c}
        \toprule
        FWHM & WMAP1 & KQ75B & KQ75M & Kp0B & KQhybrid \\
        \midrule
        0\fdg53 & 0.234 & 0.1220 & 0.1115 & 0.0812 & 0.1310 \\
        0\fdg64 & 0.286 & 0.1503 & 0.1345 & 0.1539 & 0.1604 \\
        0\fdg85 & 0.354 & 0.2608 & 0.2148 & 0.2147 & 0.2720 \\
        1\fdg28 & 0.293 & 0.3490 & 0.3167 & 0.2481 & 0.3590 \\
        1\fdg70 & 0.284 & 0.4258 & 0.3761 & 0.1360 & 0.4504 \\
        2\fdg13 & 0.248 & 0.3691 & 0.3745 & 0.1669 & 0.3822 \\
        2\fdg55 & 0.208 & 0.3205 & 0.3352 & 0.1379 & 0.3361 \\
        2\fdg98 & 0.166 & 0.2728 & 0.2684 & 0.1343 & 0.2892 \\
        3\fdg40 & 0.113 & 0.2119 & 0.2389 & 0.1023 & 0.2237 \\
        3\fdg83 & 0.081 & 0.1866 & 0.2410 & 0.0923 & 0.1963 \\
        \bottomrule
    \end{tabular}
\end{center}
\label{tab_Gfreq}
\end{table}

\subsubsection{Results of Kp0B and KQhybrid processing}
\label{subsubsec_kqp0}

\begin{figure}
    \includegraphics[angle=90, width=0.44\textwidth, trim=0 12.3cm 0.cm
0]{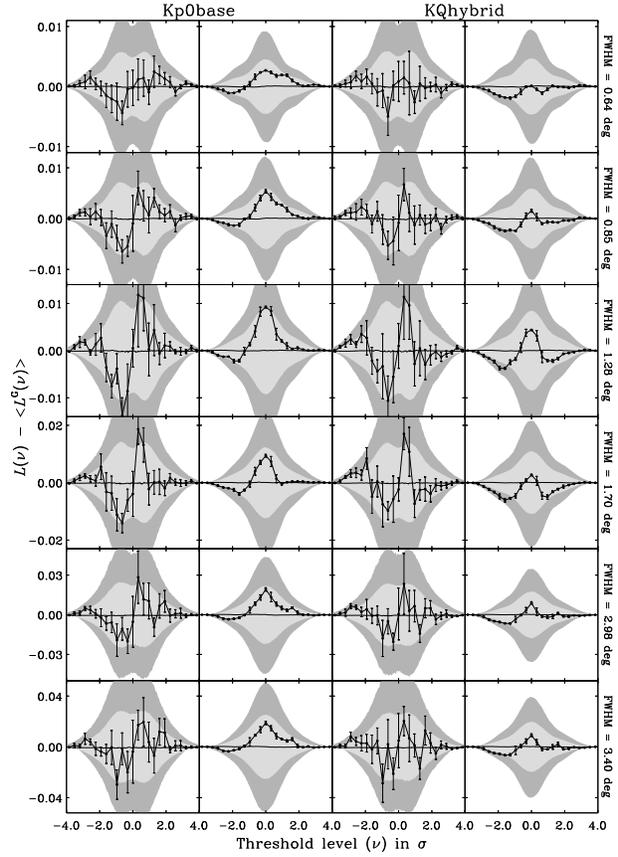} \caption{The skeleton statistics, $\Delta
\mathcal{L}(\nu)=\mathcal{L}(\nu)-\langle \mathcal{L}^{\rm G}(\nu)
\rangle$, computed from Kp0B (left two columns) and KQhybrid
(right two columns) processing on smoothing scales $\rm
FWHM=0\fdg64$, $0\fdg85$, $1\fdg28$, $1\fdg70$, $2\fdg98$ and
$3\fdg40$. The nomenclature of the elements follows the same style
as Figure \ref{fig_dske_KQ75}.} \label{fig_dske_KPQ0}
\end{figure}

We applied the one-year Kp0B mask used in
\citet{Eriksen_etal_2004} in our analysis with all other operations
remaining unchanged. We also create a new base-mask called `KQhybrid' which
excludes the same Galactic plane with KQ75B but handles the six
extended sources (Figure \ref{fig_base_mask})
identically to Kp0B. The KQhybrid mask is then included in the data
processing too as an independent test. Some of the results are shown
in Figure \ref{fig_dske_KPQ0} and the right column of Figure \ref{fig_syseff}.

In general, the $\Delta \mathcal{L}_{a}$ profiles of the Kp0B
processing are generally consistent with the previous \WMAP1 results,
although the peak-trough structure is not identical in detail.
The KQhybrid mask yields a consistent set of
results with those of KQ75B as shown in Figure
\ref{fig_dske_KPQ0}. Moreover, we have
applied the KQ75B mask to the \WMAP1 data and found that the results (dashed grey line in the left column of Figure \ref{fig_syseff}) show
a similar discrepancy from the Kp0B processed ones and consistency with
results from our 5-year VW data processing. This indicates
that modifications of the mask do significantly affect the skeleton
estimation in the \WMAP1 analysis.
Although the reason can be easily found by examining the area ratio
of the dark-grey regions in Figure \ref{fig_base_mask}, it is important
to make a separate investigation on the impact of residual Galactic
foreground and extragalactic sources since \fnl\ analysis exhibits
different responses to different types of foreground contamination
\citep{Cabella_etal_2010}. This separate analysis motivates future
skeleton studies on the effects of different Galactic foreground
templates.

It is noteworthy that the skeleton discrepancies caused by base-mask
selection indicate that residual Galactic foregrounds bias the
non-Gaussian analyses for \WMAP1 and even \WMAP3 since the Kp0 mask
was the standard temperature analysis window then and the Kp2 mask
excluded even less area around Galactic plane. This issue may have
implications on the bispectrum analysis because the additional
smoothing operation, which smears the local structures of foreground
templates, is not necessary for bispectrum estimation.

The foreground issue is also assessed as a complement to the
mask-changing analysis. We subtracted the five-year (one-year)
foreground templates from the raw maps of \WMAP1 (\WMAP5) data. The
subtracted maps are then combined and processed using the KQ75B (Kp0B)
mask and the skeleton results are depicted as the connected filled-circles (dashed black line) in the left (right) column of Figure
\ref{fig_syseff}. They are consistent with the results from the
standard foreground subtraction processing with the same corresponding
base-mask.  It is therefore confirmed that the foreground model is not
responsible for the discrepancy of the skeleton statistics as seen.

The corresponding results are listed in Table
\ref{tab_Gfreq}. It is straightforward to infer that the KQhybrid
processing results are more consistent with the corresponding KQ75B ones. The
differences of a few percent come from the 6 extended regions.
The $\chi^2$ results from the Kp0B-processing are somewhat different
to the \WMAP1 inference although the profiles of $\Delta \mathcal{L}_{a}$
are quite similar. Besides the band selection, it is most probably due
to the modified template-fitting of the Galactic foreground in
five-year data processing, as well as the better S/N level in 5-year
data.

\subsubsection{Results of median-filter processing}
\label{subsubsec_medfilt}

In order to assess the validity of the results from the KQ75B mask
analysis in which point sources are not excluded, we have applied a median
filter to those pixels located at positions in the point source mask before
smoothing, then processed the  filtered map to obtain the
skeleton statistics. Some results are plotted in the right two columns of Figure \ref{fig_dske_KQ75}, and listed in Table \ref{tab_Gfreq}.
In general, the median filtered results show good consistency with
the KQ75B results even for the first few smoothing scales,
implying that the base-mask processing is safe for skeleton analysis
on the scales considered in this work.

\begin{figure}
    \includegraphics[width=0.48\textwidth, trim=0 0 0.5cm 0]{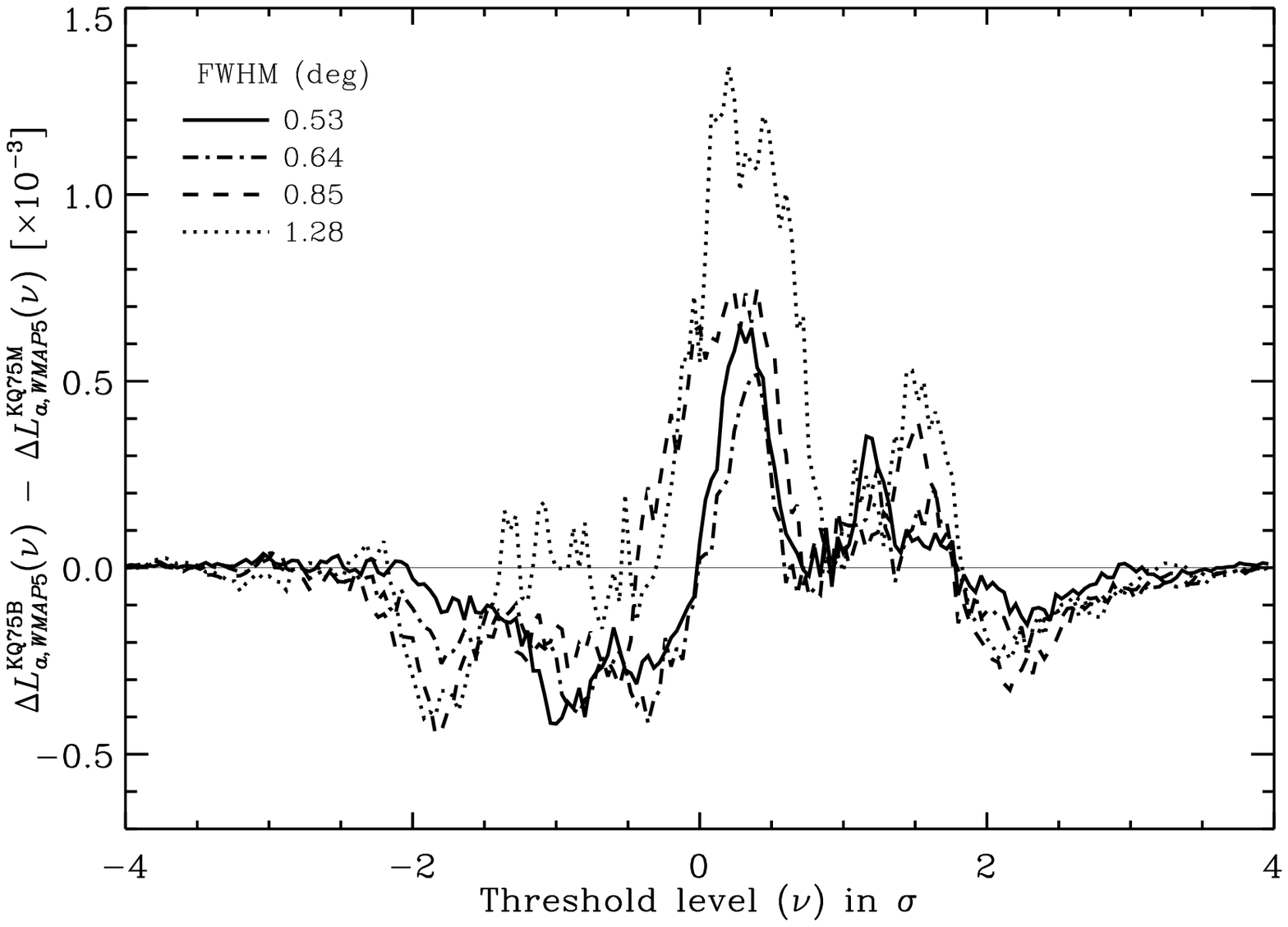}
\caption{The distribution difference between KQ75B and KQ75
median-filter processed estimator, $\Delta\mathcal{L}_{a}$ of \WMAP5
data. The cases of different smoothing scales are distinguished by
different line-styles.} \label{fig_diff_bm}
\end{figure}

Nevertheless, small visual differences suggest
further investigation into how point sources modify the skeleton
statistics and the \fnl\ estimations.
We make a comparison of the \WMAP5 $\Delta \mathcal{L}_{a}$ between
the KQ75B and median-filter processing. The differences between
them are plotted in Figure \ref{fig_diff_bm} for $\rm FWHM=0\fdg53$,
$0\fdg64$, $0\fdg85$ and $1\fdg28$ by solid, dot-dashed, dashed and dotted-lines, respectively. It is suggested that the point sources do have
asymmetric impacts on the skeleton for positive and negative temperature
thresholds - negative biasing is seen for the range $-2.0<\nu<0.0$ and
positive biasing is apparent for $0.0<\nu<2.0$.
In particular for the dotted-line, a 30\% lower depression is observed
over $0.0<\nu<0.5$. This could bias the best-fit \fnl\ value though
the bins around this range are assigned lower weights according in the
combined $\chi^2$ computation. Although the plot suggests that the
magnitude of potential biasing seems to increase with smoothing scale,
the larger smoothing still reduces sensitivity to point sources. Moreover, the profiles seen in Figure \ref{fig_diff_bm} become increasingly noise-like
within the range $\nu \in [-2.5,2.5]$ at larger smoothing scales.

\subsection{\fnl\ estimation}
\label{subsec_fnl_res}

\subsubsection{General results}
\label{subsubsec_gen_res}

Using the method introduced in Section \ref{subsec_nulltest}, the
likelihood function for \fnl\ is estimated for each
smoothing scale based on the 2500 sets of \fnl\ samples,
$\mathcal{L}^{\rm NG}(\nu,\{f_{\rm NL}\})$. We sample the parameter
within the range $f_{\rm NL} \in [-200, 400]$ with step-length
$\Delta f_{\rm NL}=2.5$. The KQ75B-processed data are utilised
from $\rm FWHM=0\fdg53$ to $3\fdg40$ with the
median-filter-processed data from $\rm FWHM=0\fdg53$ to $1\fdg28$
compared for reference. We use the cumulative estimator $\Delta
\mathcal{L}_{a}$ because it leads to 10\% more converged estimations
than the differential one according to a mock test (Section
\ref{subsec_nulltest}). Before $\chi^2$ computation, the estimator
resulting from both the observed data and simulations are rebinned to
25 bins\footnote{It has been tested that 25 is the best number for
rebinning in our analysis. More bins will make the estimator more
noisy so that the resulting likelihood is bimodal or even
multimodal, whereas less bins will make the likelihood less
converged.}.

\begin{figure}
    \includegraphics[width=0.48\textwidth,trim=0 2cm 1.5cm 1.7cm]{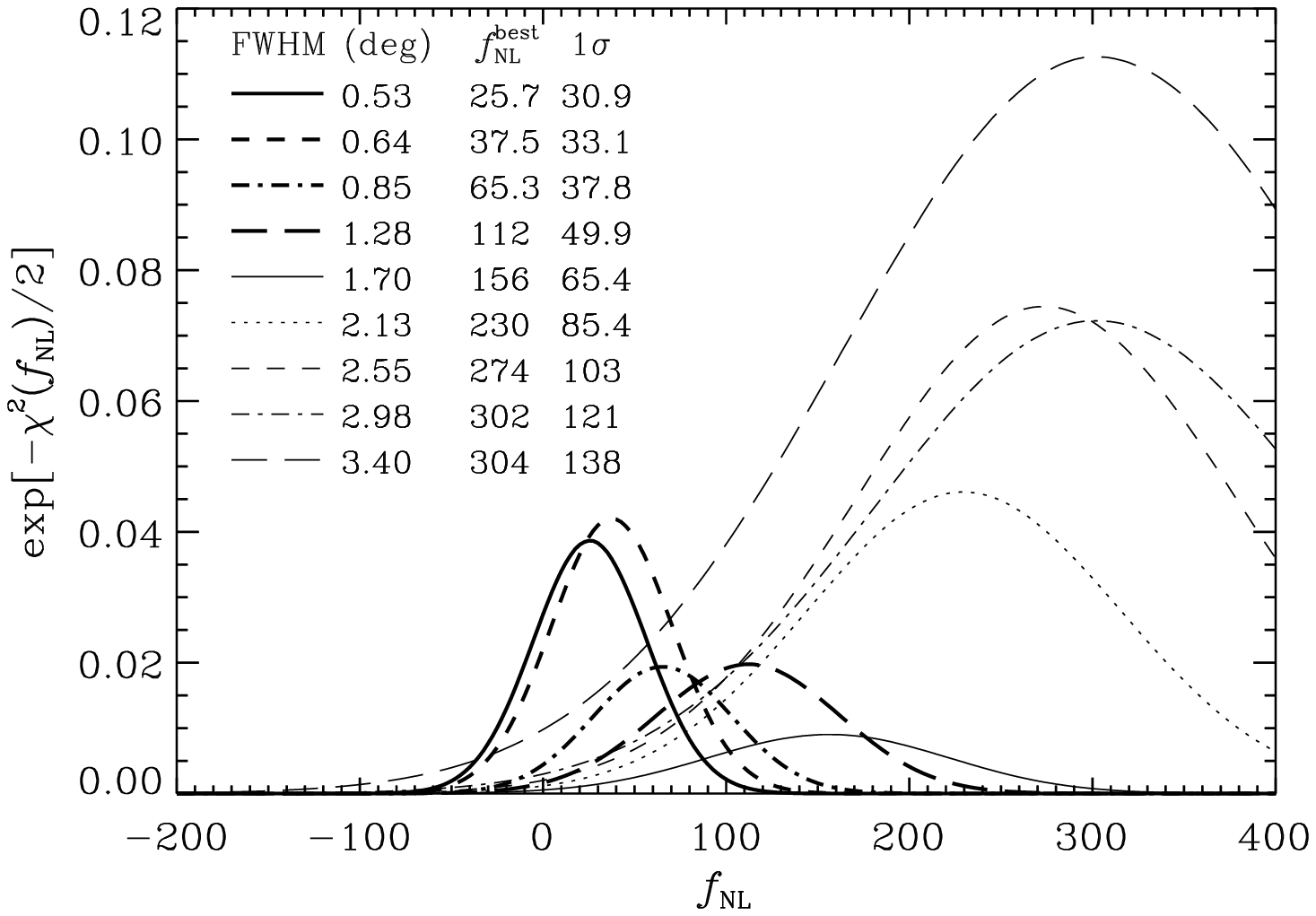}
    \includegraphics[width=0.48\textwidth,trim=0 2cm 1.5cm 0]{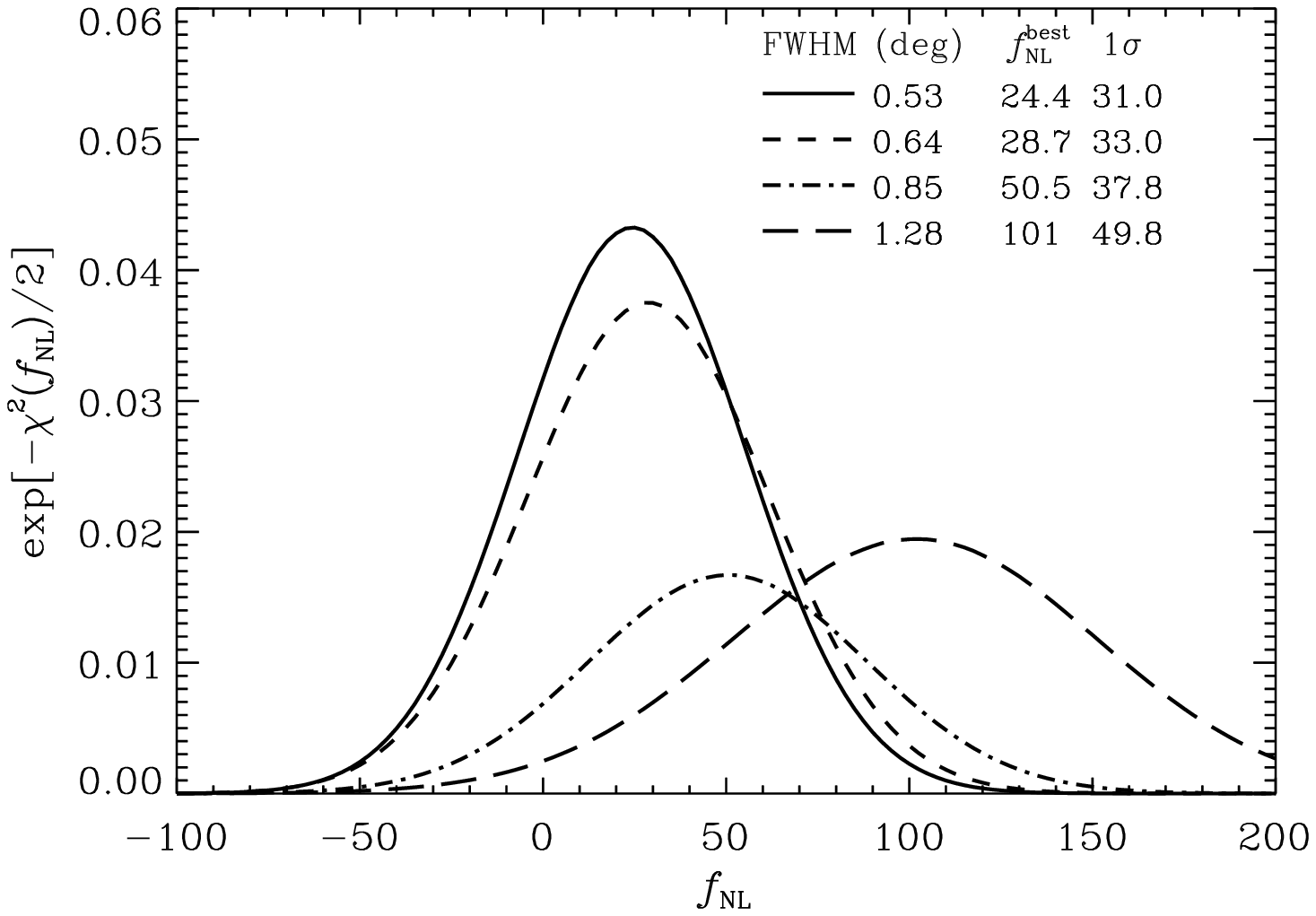}
    \includegraphics[width=0.48\textwidth,trim=0 0.5cm 1.5cm 0]{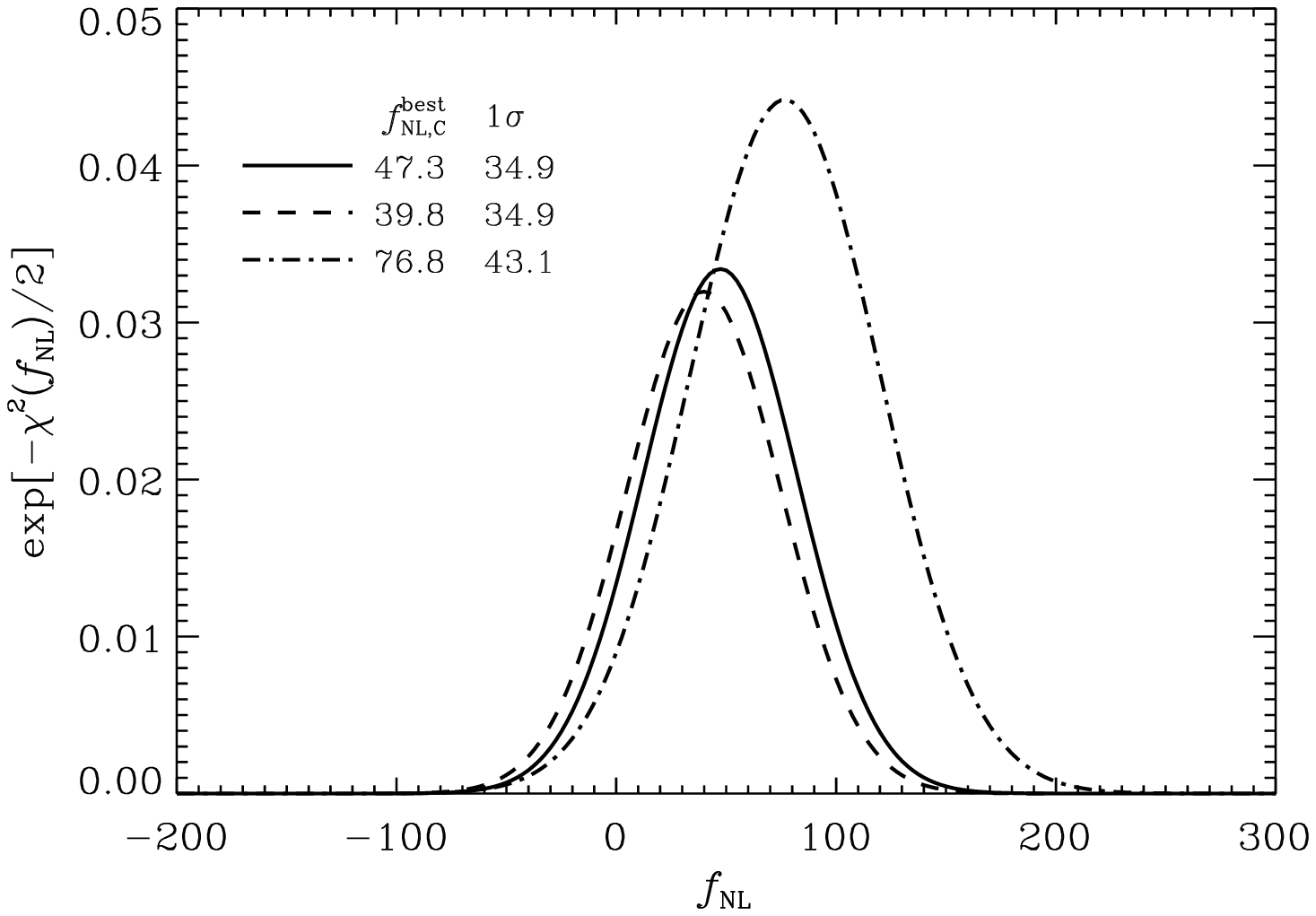}
\caption{The likelihood functions of \fnl\ from the skeleton
statistic for \WMAP5 data. \textit{Top}: The \fnl\ likelihood
functions computed by KQ75B processed
$\Delta\mathcal{L}_{a}(\nu)$ and $\mathcal{L}^{\rm
NG}_{a}(\nu,f_{\rm NL})$ on 9 different smoothing scales. The
estimator is rebinned to 25 bins before analysis. The best-fittings
and $1\sigma$ errors are obtained by fitting the likelihoods using
Gaussian functions. \textit{Middle}: Similar cases for statistic
derived by KQ75 median-filter processing on 4 smoothing scales.
\textit{Bottom}: The \fnl\ likelihood functions estimated from the
combined estimator $\Delta \mathcal{L}_{a, \rm C}(\nu,f_{\rm NG})$
and $\mathcal{L}^{\rm NG}_{a}(\nu,f_{\rm NG})$. The solid (dot-dashed)
curve shows the likelihood computed from KQ75B processed
estimator with 4 (9) FWHMs combined. The dashed curve
corresponds to the KQ75 median-filter processed likelihood.}
\label{fig_wmap_likeli}
\end{figure}

The results are shown in the top panel of Figure
\ref{fig_wmap_likeli} with each curve depicting the likelihood
(without normalisation) for each smoothing scale. The likelihood
functions are fitted by Gaussian functions so that the best-fitting
\fnl\ and the corresponding $1\sigma$ error are obtained and then
marked in the same plot. The likelihood at the highest resolution
indicates that the Gaussian hypothesis ($f_{\rm NL}=0$) is rejected
only at $0.8\sigma$-level, while it increases to $2.7\sigma$ for
$\rm FWHM=2\fdg13$. It is apparent that the best-fitting \fnl\ values show
a positive correlation with the smoothing scale, which is unexpected
since \fnl\ is scale-independent according to the local-type
non-Gaussian model and our simulations.

As discussed in Section \ref{subsubsec_medfilt}, although the
estimation is inevitably biased by the point sources or other types
of foreground, large angle smoothing renders the estimation insensitive to
those effects. We repeat the estimation using median-filtered samples
from the first four smoothing scales. As shown in the middle panel
of Figure \ref{fig_wmap_likeli}, the results are consistent in
general, and the positive correlation between $f_{\rm NL}^{\rm
best}$ and the smoothing scales is identical to the unfiltered analysis. It
is therefore suggested that the point sources contribute little to
such correlation. The $1\sigma$ errors are robust according to the
median-filter reference but the best-fit values of \fnl\ from the KQ75B
processing seem to be over-estimated by levels of $0.04\sigma$,
$0.26\sigma$, $0.39\sigma$ and $0.22\sigma$ for $\rm FWHM=0\fdg53$,
$0\fdg64$, $0\fdg85$ and $1\fdg28$, respectively.

In principle, different heights of the the likelihoods represent
variations in the goodness-of-fit if the corresponding $\chi^2$
values have the same number of degrees-of-freedom.
A higher likelihood implies the \fnl\
expectation fits the data better
and it does appear that the
likelihoods from larger-angle smoothing ($\rm FWHM=2\fdg55$,
$2\fdg98$ and $3\fdg40$) show better results than for smaller FWHMs.
However, in our analysis, we pick up only the diagonal elements of
the covariance matrix to compute the $\chi^2$. It is inappropriate to make
a theoretical interpretation of the goodness-of-fit. Consequently, the
correlation found above would be a false appearance because there
might be some bad fittings. For each FWHM, the $\chi^2$ value at the
maximum likelihood (ML) of the \emph{data} is represented as
$\chi^2_{\rm min}$. Accordingly, there are 2500 sampled
$\chi^2(\{f_{\rm NL}\}|f_{\rm NL}^{\rm ML})$ and each has a minimum
within our sampling range. We count the probability of $\chi^2_{\rm
min}(\{f_{\rm NL}\}|f_{\rm NL}^{\rm ML}) < \chi^2_{\rm min}$ to
quantify the goodness of fit for results from both the KQ75B and
median-filter processing, with a lower probability corresponding to
a better fit. The $\chi^2_{\rm min}$ values and the
probabilities are listed in Table \ref{tab_gof}. The moderate
probabilities are consistent with each other though they may be
under-estimated for the last three FWHMs. On one hand, it is
demonstrated that our skeleton statistic fits the possible
\fnl\ feature in the \WMAP5 data and our estimations are therefore
validated. On the other hand, it remains unconfirmed what the
source of the positive correlation between $f_{\rm NL}^{\rm best}$s
and smoothing scales is, and which we will return to in
Section \ref{subsubsec_Gauss_cosmic_var}.

\begin{table}
\caption{The goodness of fit, i.e., the probabilities that the
simulated samples with $\{\chi^2_{\rm min}(f_{\rm NL}|f_{\rm
NL}^{\rm ML})\} < \chi^2_{\rm min}$. $f_{\rm NL}^{\rm ML}$ and
$\chi^2_{\rm min}$ are the maximum likelihood \fnl\ of the data and
its corresponding $\chi^2$ value of each case, respectively. The
results in the case of combined data (KQ75B Comb.) are also listed.}

\centering
\begin{threeparttable}
    \begin{tabular}{c|c|c|c|c|c|c}
    \toprule
    \multirow{2}{*}{FWHM} & \multicolumn{2}{c}{KQ75B} &
    \multicolumn{2}{c}{KQ75M} & \multicolumn{2}{c}{KQ75B Comb.} \\
    & $\chi^2_{\rm min}$ & $P(\%)$ & $\chi^2_{\rm min}$ & $P(\%)$ & $\chi^2_{\rm C, min}$ & $P(\%)$
    \\
    \midrule
    $0\fdg53$ & 6.50 & 27.1 & 6.28 & 20.9 & \multirow{4}{*}{6.80} &
    \multirow{4}{*}{30.4} \\
    $0\fdg64$ & 6.34 & 25.8 & 6.57 & 22.3 & & \\
    $0\fdg85$ & 7.89 & 35.6 & 8.18 & 30.5 & & \\
    $1\fdg28$ & 7.85 & 33.5 & 7.88 & 27.3 & & \\
    \cmidrule(r){1-5}
    $1\fdg70$ & 9.42 & 40.5 & N/A & N/A & 6.24 & 28.0 \\
    $2\fdg13$ & 6.15 & 19.5 & N/A & N/A & & \\
    $2\fdg55$ & 5.20 & 12.1\tnote{a}  & N/A & N/A & & \\
    $2\fdg98$ & 5.26 & 11.7  & N/A & N/A & & \\
    $3\fdg40$ & 4.37 & 6.0  & N/A & N/A & & \\
    \bottomrule
    \end{tabular}

    \begin{tablenotes}
    \item [a] This number may be under-estimated because the underlying $\chi^2$ minima of some samples
    lay outside our \fnl\ sampling range, i.e., their corresponding $f^{\rm ML}_{\rm
    NL}>400$. Similar cases are also found for $\rm FWHM=2\fdg98$
    and $3\fdg40$.
   \end{tablenotes}
\end{threeparttable}
\label{tab_gof}
\end{table}

\subsubsection{Estimation from the combined $\Delta \mathcal{L}_{a}$}
\label{subsubsec_combination}

As presented in Section \ref{subsubsec_estimator}, the combinations
on different smoothing scales are applied separately to the rebinned
$\Delta \mathcal{L}_{a}(\nu)$ of the data and $\mathcal{L}^{\rm
NG}_{a}(\nu,f_{\rm NL})$ of the \fnl\ samples. It is verified that
such a combination still leads to an unbiased estimation of \fnl\
(Appendix \ref{sec_app_3}).

In our analysis, the first 4 and all 9 scales are combined, yielding
estimates of  $f_{\rm NL, C}=47.3\pm34.9$ and $f_{\rm NL,
C}=76.8\pm43.1$ respectively, by fitting the likelihood using a
Gaussian function. The likelihoods are shown in the bottom panel of
Figure \ref{fig_wmap_likeli} and the goodness-of-fit is also listed
in Table \ref{tab_gof}. The estimates are consistent with the
results discussed in Section \ref{subsubsec_gen_res} and the
moderate probabilities (30.4\% and 28.0\%) validate the
best-fit results.

\begin{figure}
\begin{center}
\includegraphics[width=0.48\textwidth, trim=0.5cm 0.5cm 8.5cm 0.5cm]{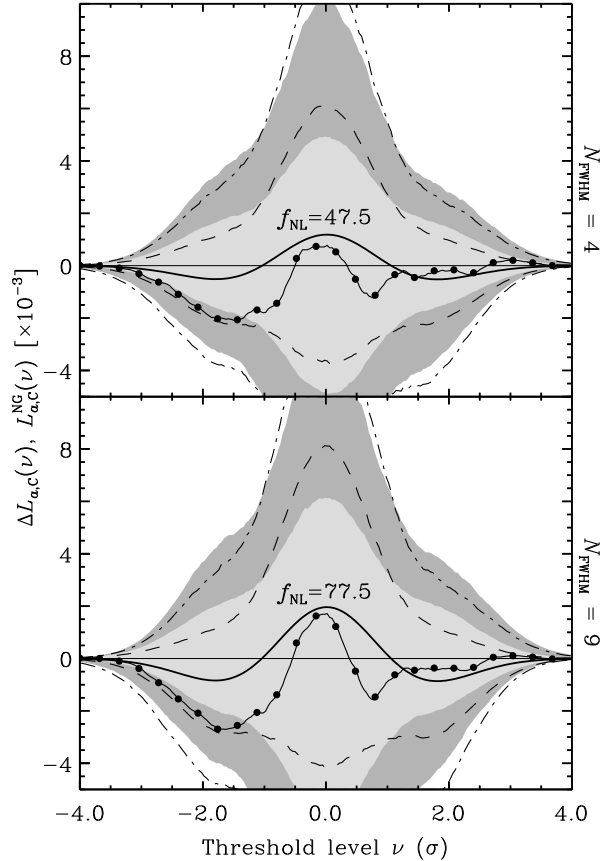}
\end{center}
\caption{The combined skeleton estimator of the data, Gaussian and
non-Gaussian predictions with $N_{\rm FWHM}=4$ (\textit{upper}) and
9 (\textit{lower}). The thin-solid curves show $\Delta \mathcal{L}_{a,\rm
C}(\nu)$ of \WMAP5 data with the filled-circles showing the
corresponding 25 bins' rebinnings. The grey bands represent the
$1\sigma$ and $2\sigma$ confidence regions of the Gaussian
predictions of $\Delta \mathcal{L}_{a,\rm C}(\nu)$. According to
2500 \fnl\ simulations, the thick-solid curves depict the expectation of
$\mathcal{L}^{\rm NG}_{a,\rm C}(\nu,f_{\rm NL}=47.5,77.5)$ for
$N_{\rm FWHM}=4$ and 9, respectively, and the dashed and
dot-dashed curves depict the corresponding $1\sigma$ and $2\sigma$
confidence boundaries.} \label{fig_dske_fnlbest}
\end{figure}

The median-filtered results are also combined over the first 4 FWHMs
and the corresponding likelihood is depicted by the dashed curve,
resulting in the estimate $f_{\rm NL, C} = 39.8\pm34.9$. The point
sources lead to an over-estimate of $f_{\rm NL, C}^{\rm best}$ at the
$0.21\sigma$-level according to this comparison. The combined
estimators, $\Delta \mathcal{L}_{a, \rm C}(\nu)$ for the KQ75B
processed data and $\mathcal{L}^{\rm NG}_{a, \rm C}(\nu,f_{\rm
NL}=0, 47.5, 77.5)$\footnote{Note that the step-length for \fnl\
sampling is 2.5 in our analysis. $f_{\rm NL}=47.5$ and $77.5$ are
the maximum likelihood values.} for the corresponding \fnl\ simulations,
are illustrated in Figure \ref{fig_dske_fnlbest} for comparison.

\subsubsection{Cosmic variance and $f_{\rm NL}^{\rm best}$}
\label{subsubsec_Gauss_cosmic_var}

It is interesting that $f_{\rm NL}^{\rm best}$s shows a monotonic
correlation with smoothing scale. The discussions above argue
against the explanation based on point sources or goodness-of-fit.
We search for this kind of correlation in our mock samples to
investigate whether cosmic variance is a possible source of such a
correlation. In order to make a comprehensive and reliable
interpretation, we pick up those Gaussian and \fnl\ samples which
show \fnl\ features at least to the same extent as the \WMAP5
data. The selection method is introduced below.

\begin{enumerate}
\renewcommand{\theenumi}{(\arabic{enumi})}

\item Gaussian samples. Similar to the \WMAP5 data, each of the 10240
Gaussian samples of $\Delta \mathcal{L}_{a}(\nu)$, is input into
\fnl-estimations on all 9 FWHMs as introduced in Section
\ref{subsubsec_estimator}. The chi-square for each FWHM,
$\chi^2_{\rm Gauss}(f_{\rm NL},\theta_{\rm FWHM})$, is obtained as
a function of smoothing scale and \fnl\ before we combine the
estimators of all 9 FWHMs to $\Delta \mathcal{L}_{\rm C}(\nu,f_{\rm
NL})$. The combined chi-square, $\chi^{2}_{\rm C, Gauss}(f_{\rm
NL})$, and likelihood are then computed by the combined estimator.
We find 3111 samples whose $\min[\chi^2_{\rm
C,Gauss}(f_{\rm NL})]$ are less than $\chi^2_{\rm C,min}$ from the
\WMAP5 data. It is believed that these samples demonstrate better
\fnl-like features (of $f^{\rm best}_{\rm NL}$) than the \WMAP5 data for
all 9 smoothing scales even though there is no non-Gaussian
component encoded in the simulations.

\item $f_{\rm NL}$ samples. For the 9-FWHM combination discussed in
Section \ref{subsubsec_combination}, samples with $\chi^{2}_{\rm C,
min}(f_{\rm NL}|f^{\rm true}_{\rm NL}=f^{\rm best}_{\rm NL,C})$ less
than the \WMAP5 $\chi^2_{\rm C,min}$ are selected from 2500 groups of
\fnl\ samples. The 701 selected samples form the \fnl\
reference for investigating the correlation between $f_{\rm NL}^{\rm
best}$s and smoothing scale.
\end{enumerate}

In 3111 selected Gaussian samples, we find 844 that
feature a monotonic correlation with smoothing scale
($\sim27.1\%$). Similarly, there are 222 \fnl\
samples from 701 showing the same behaviour  ($\sim31.7\%$).
According to our tests, they show similar properties to that
illustrated in the top panel of Figure \ref{fig_wmap_likeli} where
the maximum likelihood for large-scale smoothing is `pulled'
significantly to the non-Gaussian region. There is a considerable
probability (around 30\%) of such a correlation so
that cosmic variance is a highly probable explanation.

\section{Conclusions}
\label{sec_conclusions}

In this paper, we have studied the local-approximation to the
skeleton on a 2D sphere pixelised in the HEALPix
scheme, and refined the method of tracing the
quantity.  The statistical properties of the skeleton estimator have subsequently
been investigated using mock CMB temperature anisotropy maps.

The cubic spline interpolation method locates the skeleton knots
more accurately than the simple linear method, which makes the local
linear system more robust at the knots. This is of great importance
for finer analysis of the local system. For example, the studies on
skeleton classification \citep{Pogosyan_etal_2009}, which is
performed by analysing the eigenvalues of the linear characteristic
equation, request highly accurate estimation of such eigenvalues in
particular around the demarcation point between two types of
skeleton. Our modification provides a more reliable basis for this kind
of study. The departure of the skeleton length distribution from its
Gaussian expectation shows connections with both the sign and the
magnitude of \fnl\ so that it would yield a \fnl-likelihood
function. Based on simulated sets of CMB temperature anisotropy with
a local type  of non-Gaussian component, it has been tested that both the
differential and cumulative skeleton estimators provide unbiased and
sufficiently converged likelihood function for \fnl, but the latter
yields a likelihood 10\% more converged than the former.

The estimator was applied to the five-year \WMAP\ data release and the
results compared with not only the Gaussian predictions, but
also the results from the first-year \WMAP\ data processing by
\citet{Eriksen_etal_2004}. An \fnl-likelihood function has been
estimated by computing the $\chi^2$ on the basis of 2500 sets of \fnl\
samples. We have also investigated the goodness of fit, the impact
of the point sources and the comic variance effect on the
best-fit amplitudes of \fnl. The analysis is carried out on the V+W
combined map for various sky coverages.

The processing steps in our
analysis follow closely those of \citet{Eriksen_etal_2004} but
utilise the new five-year KQ75 mask and combined V-  and W-band data.
The smoothing scales adopted in our data processing are also identical
to those selected in \citet{Eriksen_etal_2004}. Our skeleton results
show an apparent deviation from the first-year ones. According to an
extensive series of tests,
it is the difference between the two Galactic plane regions defined by
the KQ75 and Kp0 masks that contributes mostly to the shifts.
Generally, the KQ75 mask excludes a more extended region close to the
Galactic plane than the Kp0 mask, and
this should be more conservative for temperature analysis. This kind
of deviation to the skeleton estimates implies a systematic bias in
\fnl-estimation, in other words, previous \fnl\ studies carried
out on Kp0 sky coverage (or even the less conservative KP2 mask)
may be biased by the residual Galactic foreground within the dark-grey regions
as shown in Figure \ref{fig_base_mask}. We do not exclude the pixels
located in point sources
to allow sufficient convergence of the likelihood. The impact of the point sources on the
estimator is investigated by analysing the difference to samples using
median-filtered maps. The results show that the point sources do
have an asymmetric impact on the estimator between the positive and
negative temperature thresholds on the four smallest smoothing
scales. However, the effect is less significant for larger FWHMs. The results of
our frequentist analysis show that
the \WMAP5 data are consistent with Gaussian predictions.

We have estimated the \fnl\ likelihoods on 9 smoothing levels. The
results show an unexpected positive-correlation between the
best-fit amplitudes, $f_{\rm NL}^{\rm best}$, and FWHM smoothing scales.
The peak of the likelihood function seems to be `pulled' to a highly
non-Gaussian region with the Gaussian case, $f_{\rm NL}=0$, being
`expelled' to the very tail of likelihood for some large smoothing
scales. Further investigations argue against a
point source explanation since the median-filtered data
still exhibit such a correlation. However, the presence of point sources may yield
an over-estimation of $f_{\rm NL}^{\rm best}$.

The combination of samples for the first 4 and all 9 smoothing
scales lead to the best-fit amplitudes with $1\sigma$ errors, $f_{\rm
NL}=47.3\pm34.9$ and $f_{\rm NL}=76.8\pm43.1$, respectively.
The median-filter studies
suggest that the best-fit over 4 scales may be over-estimated
at the $0.21\sigma$-level because of point sources. An
investigation has been carried out on the unexpected correlation
between $f_{\rm NL}^{\rm best}$s and smoothing scales using both
Gaussian and \fnl\ samples with a goodness-of-fit better than that for
\WMAP5.
About 30\% of them show the behaviour seen
in our analysis, so that cosmic variance may be an
appropriate explanation for this issue.

\section*{ACKNOWLEDGEMENTS}
ZH acknowledges the support by Max-Planck-Gesellschaft Chinese
Academy of Sciences Joint Doctoral Promotion Programme
(MPG-CAS-DPP), and some useful discussions with H.\,K. Eriksen, Jun
Pan, and Xi Kang. We give special thanks to St\'{e}phane Colombi for suggestions on improving the manuscript. The computations were performed at the Rechenzentrum Garching (RZG) of Max-Planck-Gesellschaft and the GPU
cluster of the cosmology group in Purple Mountain Observatory (PMO).
Some of the results in this paper have been derived using the
HEALPix \citep{Gorski_etal_2005} software and analysis package. We
acknowledge use of the Legacy Archive for Microwave Background Data
Analysis (LAMBDA) supported by the NASA Office of Space Science.

\appendix

\section{The local skeleton in HEALPix frame}
\label{sec_app_1}

\begin{figure*}
\begin{center}
    \includegraphics[width=0.96\textwidth, trim=2.5cm 0.5cm 2.0cm
21.5cm]{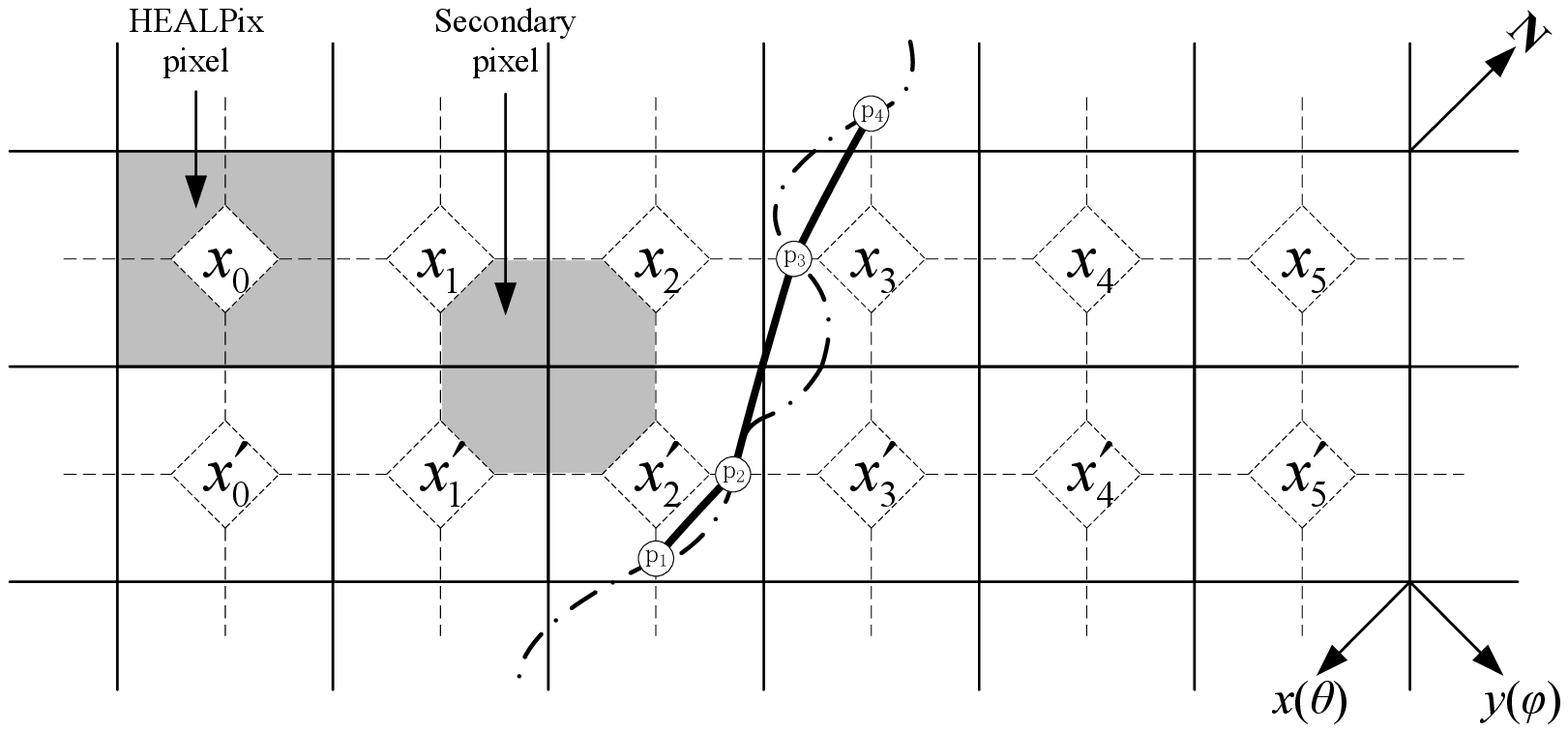}
\end{center}
\caption{Tracing the underlying contour (dot-dashed curve) on the
grid of the HEALPix pixelisation scheme. The $x_i$ mark the positions of
the primary pixels utilised in the cubic spline interpolation (Eq.
\ref{eq_cubic_S} and \ref{eq_cubic_Si}). The thick solid lines
connect the estimated intersection points ($p_i$s) between the
underlying contour and the edge of the secondary pixels by means of
a cubic spline interpolation. The definition of secondary pixels is
identical to that of \citet{Eriksen_etal_2004}. We select a coordinate
system consistent with the HEALPix convention, which is
specified by a North direction, `\textbf{N}', and two orthogonal
directions $(\theta, \phi)$ for derivative.} \label{fig_pixels}
\end{figure*}

We construct a local coordinate system on the 2D HEALPix sky map
shown in Figure \ref{fig_pixels}, where the direction to the
Galactic north-pole is depicted as `\textbf{N}'. Following the
HEALPix coordinate conventions\footnote{see `The HEALPix Primer' in
software package, version 2.10}, two orthogonal axes, $x$ and $y$ in
Eq. \ref{eq_2D_ske}, are set to be aligned with the polar-angle
$\theta$ and azimuth $\phi$ axes, respectively.

Our starting point is identical with that of
\citet{Eriksen_etal_2004} in that we determine a pair of vertices on the
edge of the pre-constructed secondary pixels on $\mathcal{S}$, with
one vertex value lower but the other higher than zero (canceling
vertices, hereafter). It is suggested that the underlying skeleton
crosses over those edges connecting pairs of canceling vertices.
Figure \ref{fig_pixels} illustrates an exaggerated version of this
process. Interpolation is then necessary to determine the
positions of the intersections on edges (skeleton knots, hereafter).
Linear interpolation has been
adopted previously \citep{Novikov_etal_1999, Shandarin_etal_2002,
Eriksen_etal_2004}, since it has been widely employed in
morphological studies on both the CMB and large scale structures, eg.
the length and genus quantities of MFs, which are related to the
contour lines of the random fluctuation field and its derivatives.
However, the accuracy of an interpolation method is limited by the
topological properties of the random field and the pixel size of the
corresponding realisation. Linear interpolation is accurate
enough for an analysis of the MFs at current observational resolutions
($N_{\rm side} = 512, 1024$), however, it is inadequate to
provide precise positions of `knots' on the skeleton map in Eq.
\ref{eq_CMB_ske} as a higher-order (cubic) random field. This may
introduce not only bias in to the statistics of the skeleton length for a specific
realisation, but also could result in a false determination of the eigenvalue
of the local linear system, in particular around the demarcation
point between the two types of skeletons considered for studies of skeleton
classifications \citep{Pogosyan_etal_2009}. Given the cubic
nature of the skeleton field, we therefore apply a cubic spline
interpolation in this analysis, as introduced in the following
text in detail. A comparison between the linear and cubic spline
strategy is presented in Appendix \ref{sec_app_2}.

Once a pair of canceling vertices has been found, e.g., $x_{2}$ and
$x_{3}$ in Figure \ref{fig_pixels}, the 6 pixels are then picked up
with canceling vertices in the middle, as $x_{0}$, $x_{1}$, ...,
$x_{5}$. The connection lines of the 6-pixel centres must cross over
the pairs of opposite sides of the quadrangular pixels and be
parallel with the connection line of canceling vertices (e.g.,
$\overline{x_{2}x_{3}}$). The values of these 6 pixels (vertices of
secondary pixels), $y_{i} = \mathcal{S}(x_{i}) (i=0,1,2,3,4,5)$, are
utilised to find the spline functions along the connection lines,
\begin{equation}
\mathcal{S}(x) = \left\{\begin{matrix}
    S_0(x) & x \in [x_{0}, x_{1}] \\
    S_1(x) & x \in [x_{1}, x_{2}] \\
    \vdots & \vdots \\
    S_4(x) & x \in [x_{4}, x_{5}]
\end{matrix}\right.,
\label{eq_cubic_S}
\end{equation}
where each $S_i$ is the piecewise cubic polynomial between the
pixel-centres
\begin{eqnarray}
S_i(x) & = & \frac{z_{i+1} (x-x_i)^3 + z_i (x_{i+1}-x)^3}{6h_i}
       \nonumber \\
       & & \mbox{} + \left(\frac{y_{i+1}}{h_i} - \frac{h_i}{6} z_{i+1}\right)(x-x_i)
       \nonumber \\
       & & \mbox{} + \left(\frac{y_{i}}{h_i} - \frac{h_i}{6} z_i\right) (x_{i+1}-x).
\label{eq_cubic_Si}
\end{eqnarray}
$h_i$ is equal to $|\vecx_{i+1}-\vecx_{i}|$ corresponding to the
radial distance of the two pixel-centres. The coefficients
$\{z_{i}\}$ can be obtained by solving the linear system
\begin{equation}
\begin{split}
h_{i-1}z_{i-1} + & 2(h_{i-1} + h_i)z_i + h_iz_{i+1} = \\
& 6\left(\frac{y_{i+1}-y_i}{h_i} -
\frac{y_i-y_{i-1}}{h_{i-1}}\right), i=1,2,3,4 \\
& z_0 = z_5 = 0.
\end{split}
\end{equation}
where $y_i$ corresponds to the pixel value at $x_i$, i.e., the skeleton
value, $\mathcal{S}(x_i)$, in this work.

Note that this 6-point system on the sphere has been approximated by a
1D straight line since the pixel-size in our analysis is so small
($N_{\rm side} = 1024$, $\theta_{\rm pix}\sim3.44\arcmin$). In fact,
we only need $S_2(x)$ to determine the locations of the knots, e.g.,
$p_3$ in Figure \ref{fig_pixels}, by solving the cubic equation
\begin{equation}
S_2(x) = 0.
\end{equation}
There is one and only one real root, $x_{k}$, satisfying the
condition $x_2<x_{k}<x_3$. Then the vector of the underlying knot
can be obtained as
\begin{equation}
\vecx_{k} = \frac{x_3-x_k}{x_3-x_2}\vecx_2 +
\frac{x_k-x_2}{x_3-x_2}\vecx_3,
\end{equation}
and the corresponding temperature value at $\vecx_{k}$ is
\begin{equation}
T_k = \frac{x_3-x_k}{x_3-x_2}T_2 + \frac{x_k-x_2}{x_3-x_2}T_3.
\label{eq_T_intpl}
\end{equation}
According to Figure \ref{fig_pixels}, after determining the vector
of $p_2$ and $p_3$ (i.e., $\vecx_{k2}$ and $\vecx_{k3}$), the
skeleton length within the secondary pixel, $x_2x_3x'_3x'_2$, can be
estimated by the dot-product of these two vectors,
\begin{equation}
\delta L(T_s) =
\arccos\left(\frac{\vecx_{k2}}{|\vecx_{k2}|}\cdot\frac{\vecx_{k3}}{|\vecx_{k3}|}\right).
\end{equation}
The corresponding temperature value of this piece of skeleton
length, $T_s$, is approximately the simple average of $T_{k2}$ and
$T_{k3}$.

It is always the case that the four edges of one secondary pixel
are connecting canceling vertices. Most of these cases indicate
a stationary point (maxima or minima or saddle point) within this
secondary pixel, implying two skeletons cross inside. There are
still a few exceptions but they will become very rare due to the
small pixel-size and the smoothing applied afterwards. We therefore
make the same assumption as in \citet{Eriksen_etal_2004} that all of
the cases indicate a pair of skeletons crossing over each other.
The possible deviation from the length distribution is totally
negligible according to various tests.

\section{Comparison between Linear and cubic spline interpolation for skeleton analysis}
\label{sec_app_2}

\begin{figure}
    \includegraphics[width=0.48\textwidth, trim=0 0 8cm 0]{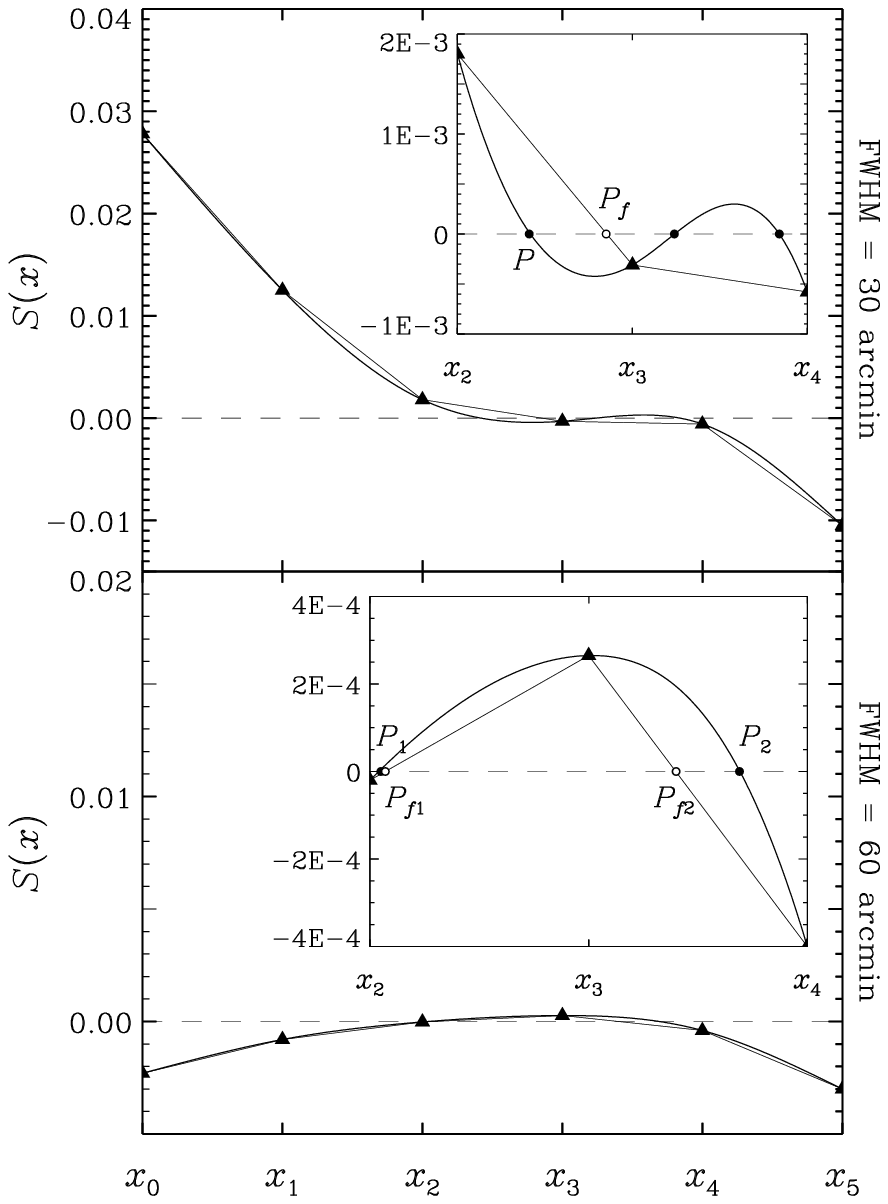}
\caption{The pieces of profile of $\mathcal{S}$ within a
six-pixel-array reproduced by linear and cubic spline interpolation
from \emph{one} Gaussian simulation. The centres of the six pixels
are identified by $x_0$, $x_1$, ..., $x_5$ where the values of
$\mathcal{S}$ are marked by filled triangles. The positions of the
skeleton knots estimated by cubic splines (linear lines) are located
by the filled (open) circles. The small plots inside zoom the
curves within $x_2$ and $x_4$. The pixel index (position) of $x_i$
is selected to be the \emph{same} for $\rm FWHM=30\arcmin$ and
$60\arcmin$.} \label{fig_lin_cub}
\end{figure}

On a pixelised 2D random field, the key step in tracing the local
skeleton is to locate the skeleton knot which is always within the
line connecting the centres of the two canceling neighbouring pixels
(one edge of the secondary pixel), and whose position is conventionally
estimated by linear interpolation, since the skeleton
realisation $\mathcal{S}$ can be considered as a linear function
along the line connecting just a few pixels at a very high
resolution-level. This is an approximation that makes things easier
to handle, especially for the HEALPix pixelization scheme. However,
the skeleton is actually a cubic function, so that it is necessary to test whether
linear interpolation is sufficient for its computation. In this appendix, we
investigate the linear properties at the skeleton knots derived by
linear and cubic spline interpolation methods.

\begin{figure}
    \includegraphics[width=0.48\textwidth, trim=0 0 0.5cm 0.0cm]{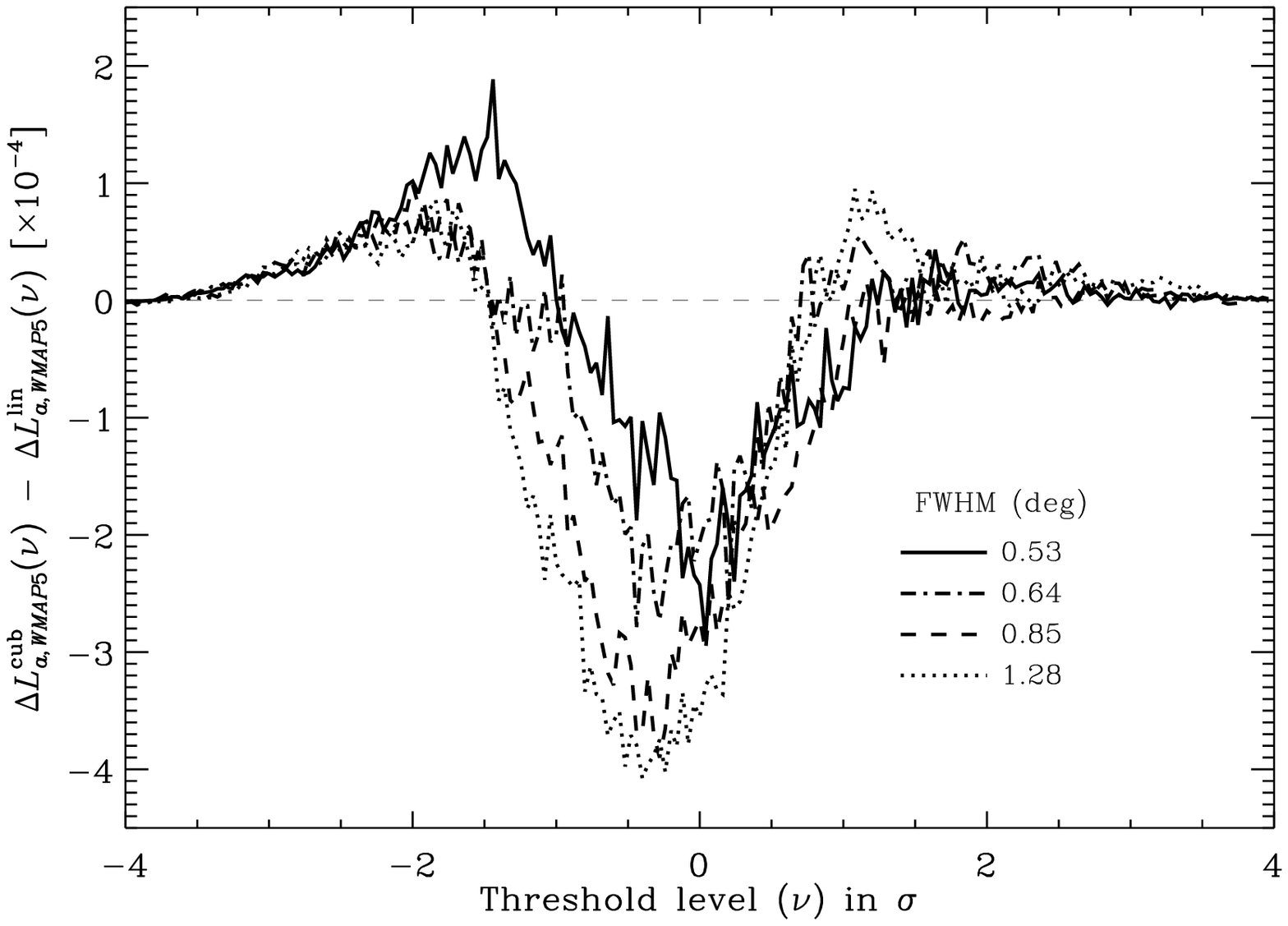}
\caption{The difference of $\Delta \mathcal{L}_{a}$ of \WMAP5 data
between the cubic spline (cub) and linear (lin) interpolation
processing. The cases of different smoothing scales are
distinguished by different line-styles.} \label{fig_diff_lincub}
\end{figure}

The characteristic equation (Eq. \ref{eq_det}) for
the 2D random field must be satisfied at the skeleton knots. It can be
reexpressed for a CMB temperature field as
\begin{equation}
\begin{pmatrix} T_{;\theta \theta} & T_{;\theta
\phi} \\ T_{;\phi \theta} & T_{;\phi \phi} \end{pmatrix}
\begin{pmatrix} T_{;\theta} \\ T_{;\phi} \end{pmatrix} = \lambda
\begin{pmatrix} T_{;\theta} \\ T_{;\phi} \end{pmatrix}.
\end{equation}
We define
\begin{equation*}
r_1 \equiv \frac{T_{;\theta \theta}T_{;\theta}+T_{;\theta
\phi}T_{;\phi}}{T_{;\theta}}, \ r_2 \equiv \frac{T_{;\phi
\theta}T_{;\theta}+T_{;\phi \phi}T_{;\phi}}{T_{;\phi}},
\end{equation*}
and $\lambda$ should satisfy the following
\begin{equation}
\begin{vmatrix} T_{;\theta \theta}-\lambda & T_{;\theta
\phi} \\ T_{;\phi \theta} & T_{;\phi \phi}-\lambda \end{vmatrix} = 0
\label{eq_det_eigenvalue}
\end{equation}
with two real roots $\lambda_1$ and $\lambda_2$ ($\lambda_1 \ge
\lambda_2$). In principle, $r_1$ should be equal to $r_2$ and also
equal to $\lambda_1$ or $\lambda_2$ along the underlying skeleton.
However, in practice, we have to investigate such a property at the
position of the estimated skeleton knots on the pixelised sphere where
$r_1$ and $r_2$ are not exactly equal because of estimation
errors. The first and second derivatives there can be obtained safely
by linear interpolation as in  Eq. \ref{eq_T_intpl}. We define a new
quantity $r \equiv (r_1+r_2)/2$. The numerical robustness of the
equivalence between $r$ and $\lambda$ indicates the quality of the
estimation method.

In this test, we pick up two six-pixel-arrays ($x_0, x_1, ..., x_5$)
from one simulated Gaussian realisation (resolution parameter
$N_{\rm side}=1024$) smoothed by Gaussian beams with $\rm
FWHM=30\arcmin$ and $60\arcmin$.
The pixel location of the two arrays are exactly the same with each
other.
The corresponding
values of $\mathcal{S}$ in Eq. \ref{eq_CMB_ske} are marked by filled
triangles in Figure \ref{fig_lin_cub}. For the case of $\rm
FWHM=30\arcmin$, $x_2 x_3$ is a pair of canceling pixels and $P_f$
point ($P$ point) is the estimated skeleton knot determined by a linear (cubic
spline) interpolation method. The linear properties at the two
points are quantified as {\setlength\arraycolsep{1pt}
\begin{eqnarray*}
P_f: \ r_1 & = & -0.6887, \ r_2 = -0.7466, \ r = -0.7174 \\
\lambda_1 & = & -0.6422, \ \lambda_2 = -0.7221 \\
P: \ r_1 & = & -0.7099, \ r_2 = -0.7116, \ r = \bf{-0.7108} \\
\lambda_1 & = & -0.6125, \ \lambda_2 = \bf{-0.7107} \ \rm(secondary
\ skeleton)
\end{eqnarray*}}
It is shown in this example that the cubic spline interpolation
leads to a more accurate location of the skeleton knots, and the
distribution of skeleton length therein. Note that there are two
suspicious skeleton knots within $x_3$ and $x_4$ in this case but
they would not be involved in analysis since $x_3$ and $x_4$ are not
canceling pixels. It is also noteworthy that the point $P$ belongs
to a piece of the first-type secondary skeleton according to the
classification in \citet{Pogosyan_etal_2009}. The robust equivalence
between $r$ and the eigenvalue indicates accurate and unbiased
classification, in particular around the underlying demarcation point
between two types of skeleton where the two eigenvalues are quite
close to each other. The cases for $\rm FWHM=60\arcmin$ are listed
below
{\setlength\arraycolsep{1pt}
\begin{eqnarray*}
P_{f1}: \ r_1 & = & 0.0428, \ r_2 = -0.4198, \ r = -0.2313 \\
\lambda_1 & = & -0.2114, \ \lambda_2 = -0.8062 \\
P_{1}: \ r_1 & = & -0.1916, \ r_2 = -0.2425, \ r = \bf{-0.2170} \\
\lambda_1 & = & {\bf{-0.2108}}, \ \lambda_2 = -0.8060 \ \rm(primary
\ skeleton)
\end{eqnarray*}}
{\setlength\arraycolsep{1pt}
\begin{eqnarray*}
P_{f2}: \ r_1 & = & -0.9264, \ r_2 = -0.7597, \ r = -0.8430 \\
\lambda_1 & = & -0.2230, \ \lambda_2 = -0.8091 \\
P_{2}: \ r_1 & = & -0.7981, \ r_2 = -0.8004, \ r = \bf{-0.7993} \\
\lambda_1 & = & -0.2588, \ \lambda_2 = \bf{-0.7996} \ \rm(secondary
\ skeleton)
\end{eqnarray*}}

For the difference between the five-year and one-year skeleton
processing, we must investigate the impact of method selection on
the results. Given the KQ75B processed data and Gaussian
simulations, we carry out the skeleton analysis following the steps
described in Section \ref{subsec_analysis} but utilising linear
interpolation to locate the skeleton knots. The resulting length
departure of the data is then obtained
\begin{equation}
\Delta \mathcal{L}^{\rm lin}_{a,\emph{WMAP}\rm5} = \mathcal{L}^{\rm
lin}_{a,\emph{WMAP}\rm5} - \langle\mathcal{L}^{\rm G,lin}_{a}\rangle
\end{equation}
and the differences between the cubic spline and linear results are
plotted in Figure \ref{fig_diff_lincub} for $\rm FWHM=0\fdg53$,
$0\fdg64$, $0\fdg85$ and $1\fdg28$. It is noteworthy that the
magnitude of such a difference contributes less than 10\% to
the discrepancy between the \WMAP5 and \WMAP1 skeleton length
distribution profile. However, the structure shown in Figure
\ref{fig_diff_lincub} suggests that the linear method would lead to
an over-enhanced peak and over-depressed trough, which for the
positive-\fnl\ structure of $\Delta \mathcal{L}_{a}$ suggested by the data
may bias the best-fitting value of \fnl.

\section{Test of the likelihoods from the combined estimator}
\label{sec_app_3}

\begin{figure}
\begin{center}
\includegraphics[width=0.48\textwidth, trim=1cm 0.7cm 2cm 1.7cm]{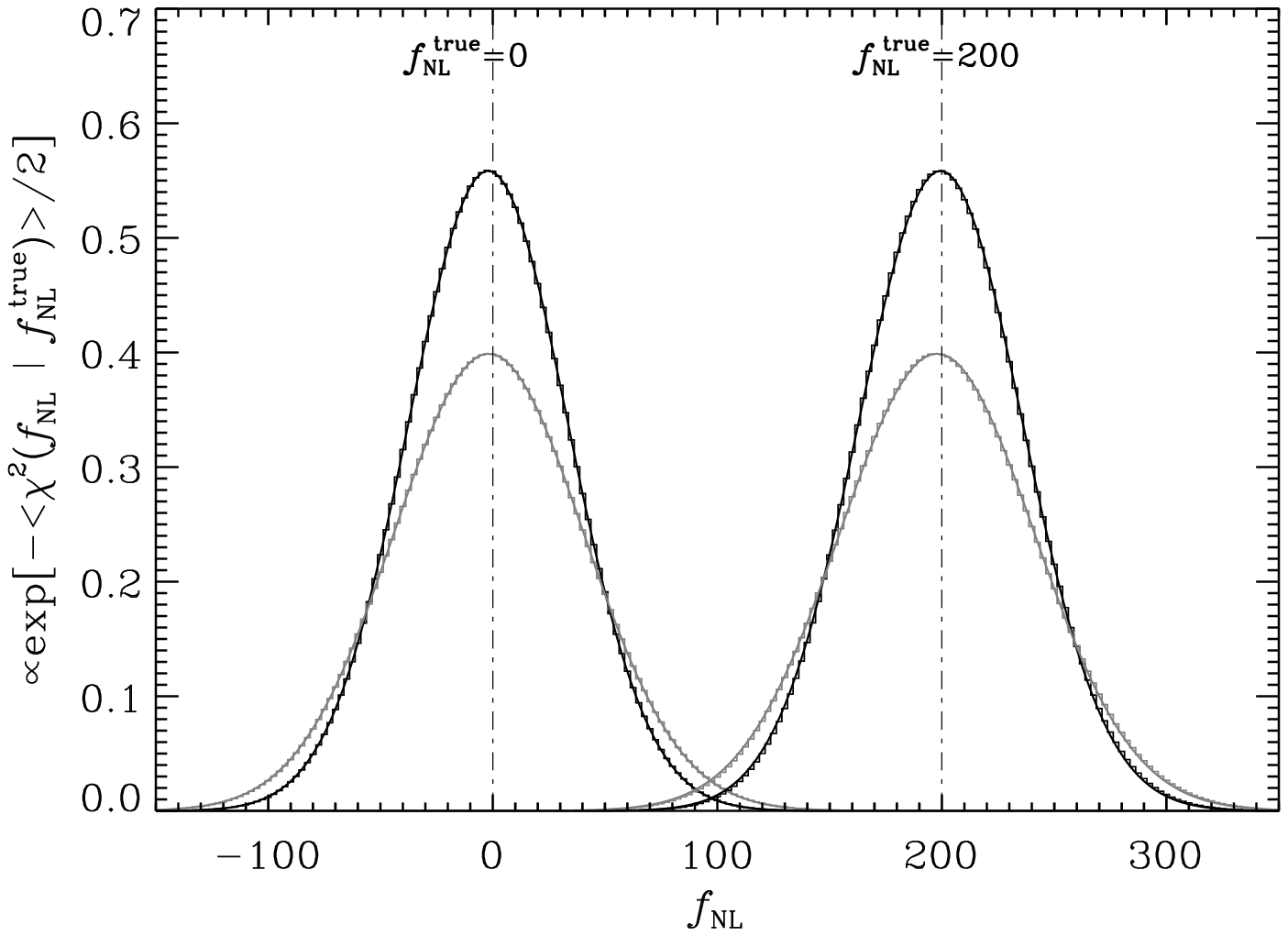}
\end{center}
\caption{$\exp\left[-\frac{1}{2N}\sum^{N}_{j=1}\chi^2(f^{j}_{\rm
NL}|f^{\rm true}_{\rm NL})\right]$, the effective likelihood
functions computed by the combined accumulative estimator
$\mathcal{L}^{\rm NG}_{a,\rm C}$ from KQ75B processed noisy
simulations with input parameter $f_{\rm NL}^{\rm true}=0,200$. The
functions are renormalised by different factors for visual
convenience and are shown by histograms. The Gaussian-fitting
functions are depicted by solid curves. The higher, narrower (lower,
wider) histograms and curves correspond to the combinations with
$N_{\rm FWHM}=4$ ($9$).} \label{fig_likeli_comb}
\end{figure}

In this section we test for the presence of bias in our combined estimator.
Given simulated noisy realisations from the KQ75B
processing and the predetermined expectation $\langle\mathcal{L}^{\rm
NG}_{\rm C}(\nu,f_{\rm NL})\rangle$, we randomly pick up $N=250$
sets of \fnl-samples, $\mathcal{L}^{\rm NG}_{\rm C}(\nu,f_{\rm
NL}^{j})$ ($j=1,2,...,250$) with $N_{\rm FWHM} = 4$ and $9$, to form
the conditional $\chi^2$ functions
\begin{equation}
\chi^2_{\rm C}(f_{\rm NL}^{j}|f_{\rm NL}^{\rm true})=\sum_{\nu}
\left\{ \frac{\mathcal{L}^{\rm NG}_{\rm C}(\nu,f_{\rm
NL}^{j})-\langle \mathcal{L}^{\rm NG}_{\rm C}(\nu,f_{\rm NL}^{\rm
true}) \rangle}{\sigma[\mathcal{L}^{\rm NG}_{\rm C}(\nu,f_{\rm
NL}^{\rm true})]} \right\}^2,
\end{equation}
and the effective likelihood function for each sample,
\begin{equation}
\mathscr{L}_{\rm C}(f_{\rm NL}|f_{\rm NL}^{\rm true}) \propto
\exp\left[-\frac{1}{2N}\sum^{N}_{j=1}\chi^2_{\rm C}(f_{\rm
NL}^{j}|f_{\rm NL}^{\rm true})\right].
\end{equation}

We plot $\mathscr{L}_{\rm C}(f_{\rm NL}|f_{\rm NL}^{\rm true})$ as
histograms for two given $f_{\rm NL}^{\rm true}$ values (0 and 200)
in Figure \ref{fig_likeli_comb} for $N_{\rm FWHM}=4$ and $9$,
noticing that the sampling width $\Delta f_{\rm NL}$ is 2.5. Again,
the likelihoods are perfectly fitted by Gaussian functions with the
parameters listed in Table \ref{tab_cmbntest}. Despite the noise
contribution and sky-cut, it is demonstrated that the
inverse-variance-combination still leads to an unbiased skeleton
estimator for \fnl.

\begin{table}
\caption{The Maximum-Likelihood ($f_{\rm NL}^{\rm ML}$),
best-fitting ($f_{\rm NL}^{\rm best}$) values and $1\sigma$ error
from the likelihood $\mathscr{L}_{\rm C}(f_{\rm NL}|f_{\rm NL}^{\rm
true})$ (Figure \ref{fig_likeli_comb}) computed from the combined
estimator derived from $N=250$ KQ75B processed noisy simulations
with given parameter $f_{\rm NL}^{\rm true}=0,200$.}
\begin{center}
    \begin{tabular}{c|c|c|c|c}
        \toprule
        $N_{\rm FWHM}$ & $f_{\rm NL}^{\rm true}$ & $f_{\rm NL}^{\rm ML}$
        & $f_{\rm NL}^{\rm best}$ & $\sigma_{f_{\rm NL}}$ \\
        4 & 0.0 & -2.5 & -2.0 & 35.5 \\
        9 & 0.0 & -2.5 & -1.7 & 42.3 \\
        4 & 200.0 & 200.0 & 199.4 & 36.5 \\
        9 & 200.0 & 197.5 & 197.8 & 43.4 \\
        \bottomrule
    \end{tabular}
\end{center}
\label{tab_cmbntest}
\end{table}

\label{lastpage}
\end{document}